\documentclass[aps,
               pra,
               twocolumn,
               nofootinbib,
               floatfix,
               showpacs
              ]{revtex4}

\usepackage{amsmath, amssymb}
\usepackage{graphics}
\usepackage{pstricks}
\usepackage{graphicx}
\usepackage{color}
\definecolor{darkblue}{rgb}{0.0,0.0,0.4}
\definecolor{darkgreen}{rgb}{0.0,0.4,0.0}
\definecolor{darkred}{rgb}{0.6,0.0,0.0}
\usepackage[bookmarksopen]{hyperref}
\hypersetup{colorlinks,linkcolor=darkgreen,citecolor=darkred,urlcolor=darkblue}

\renewcommand{\i}{\mathrm{i}}

\newcommand{\id}{\mathrm{d}}
\renewcommand{\vector}[1]{\mathbf{#1}}

\newcommand{\cred}[1]{{\color{red}{#1}}}

\newcommand{\weg}[1]{{\cred{}}}

\newcommand{\del}{\partial}

\newcommand{\Eq}[1]{Eq.~(\ref{eq:#1})}

\newcommand{\Fig}[1]{Fig.~\ref{fig:#1}}

\newcommand{\Sect}[1]{Sect.~\ref{sec:#1}}
\newcommand{\Subsect}[1]{Sect.~\ref{subsec:#1}}

\newcommand{\subsect}[1]{\ref{subsec:#1}}

\newcommand{\App}[1]{App.~\ref{app:#1}}

\begin{document}

\title{Critical Dynamics of a Two-dimensional Superfluid near a Non-Thermal Fixed Point}

\author{Jan Schole}
\author{Boris Nowak}
\author{Thomas~Gasenzer}
\email{t.gasenzer@uni-heidelberg.de}
\affiliation{Institut f\"ur Theoretische Physik,
             Ruprecht-Karls-Universit\"at Heidelberg,
             Philosophenweg~16,
             69120~Heidelberg, Germany}
\affiliation{ExtreMe Matter Institute EMMI,
             GSI Helmholtzzentrum f\"ur Schwerionenforschung GmbH, 
             Planckstra\ss e~1, 
             64291~Darmstadt, Germany} 

\date{\today}

\begin{abstract}
Critical dynamics of an ultracold Bose gas far from equilibrium is studied in two spatial dimensions. Superfluid turbulence is created by quenching the equilibrium state close to zero temperature. Instead of immediately re-thermalizing, the system approaches a meta-stable transient state, characterized as a non-thermal fixed point. A focus is set on the vortex density and vortex-antivortex correlations which characterize the evolution towards the non-thermal fixed point and the departure to final (quasi-)condensation. Two distinct power-law regimes in the vortex-density decay are found and discussed in terms of a vortex unbinding process and a kinetic description of vortex scattering. A possible relation to decaying turbulence in classical fluids is pointed out. By comparing the results to equilibrium studies of a two-dimensional Bose gas, an intuitive understanding of the location of the non-thermal fixed point in a reduced phase space is developed. 
\end{abstract}


\maketitle

\section{Introduction}
\label{sec:intro}
Generically, the properties of an interacting many-body system far from equilibrium can change violently in time. This is not the case, however, when the system reaches a non-thermal fixed point. In the vicinity of such a fixed point the dynamics is expected to slow down and correlation functions to exhibit universal scaling behavior. Such transient states have been intensively studied in the context of classical turbulence~\cite{Richardson1920a,Kolmogorov1941a,Obukhov1941a,Frisch1995a}.
Similar phenomena appear in degenerate quantum many-body systems, e.g., in superfluid helium or dilute ultracold quantum gases where superfluid or quantum turbulence (QT) has been discussed in great detail \cite{Nore1997a,Vinen2002a,Araki2002a,Kobayashi2005a,Proment2010a,Henn2009a,Seman2011a,Neely2012a}.
More general types of turbulence, so called wave turbulence have been studied  \cite{Zakharov1992a,Nazarenko2011a}, mainly in the framework of kinetic theory. 
Recent applications include far-from-equilibrium quantum systems, such as dilute ultracold Bose gases \cite{Levich1978a,Kagan1992a,Kagan1994a,Kagan1997c,Berloff2002a,Svistunov2001a,Kozik2009a,Scheppach:2009wu,Nowak:2010tm,Nowak:2011sk,Schmidt:2012kw}, the inflating and reheating early universe \cite{Micha:2002ey,Berges:2008wm,Berges:2008sr,Berges:2010ez,Gasenzer:2011by}, and quark-gluon matter created in heavy-ion collisions \cite{Arnold:2005ef, Berges:2008mr, Carrington:2010sz, Fukushima:2011nq, Fukushima:2011ca,Blaizot:2011xf,Berges:2012us}.
Thereby, an extension of kinetic wave-turbulence by non-perturbative quantum-field-theory methods lead to the notion of a non-thermal fixed point \cite{Berges:2008wm,Berges:2008sr,Gasenzer:2011by,Carrington:2010sz}, in analogy to fixed points describing equilibrium as well as dynamical critical phenomena \cite{Hohenberg1977a}.

It was demonstrated in Refs.~\cite{Nowak:2010tm,Nowak:2011sk} that in a 2- or 3-dimensional superfluid Bose gas, such a non-thermal fixed point is realized by a state with a dilute random distribution of vortices or vortex lines, respectively, of both, positive and negative circulation. 
This gave the relation to QT \cite{Nore1997a,Vinen2002a,Araki2002a,Kobayashi2005a,Proment2010a,Nazarenko2011a} as well as a link between weak wave turbulence described by kinetic theory and topological excitations in non-linear wave systems. 

    \begin{figure}[!t]
   \includegraphics[width=0.28\textwidth,angle=-90]{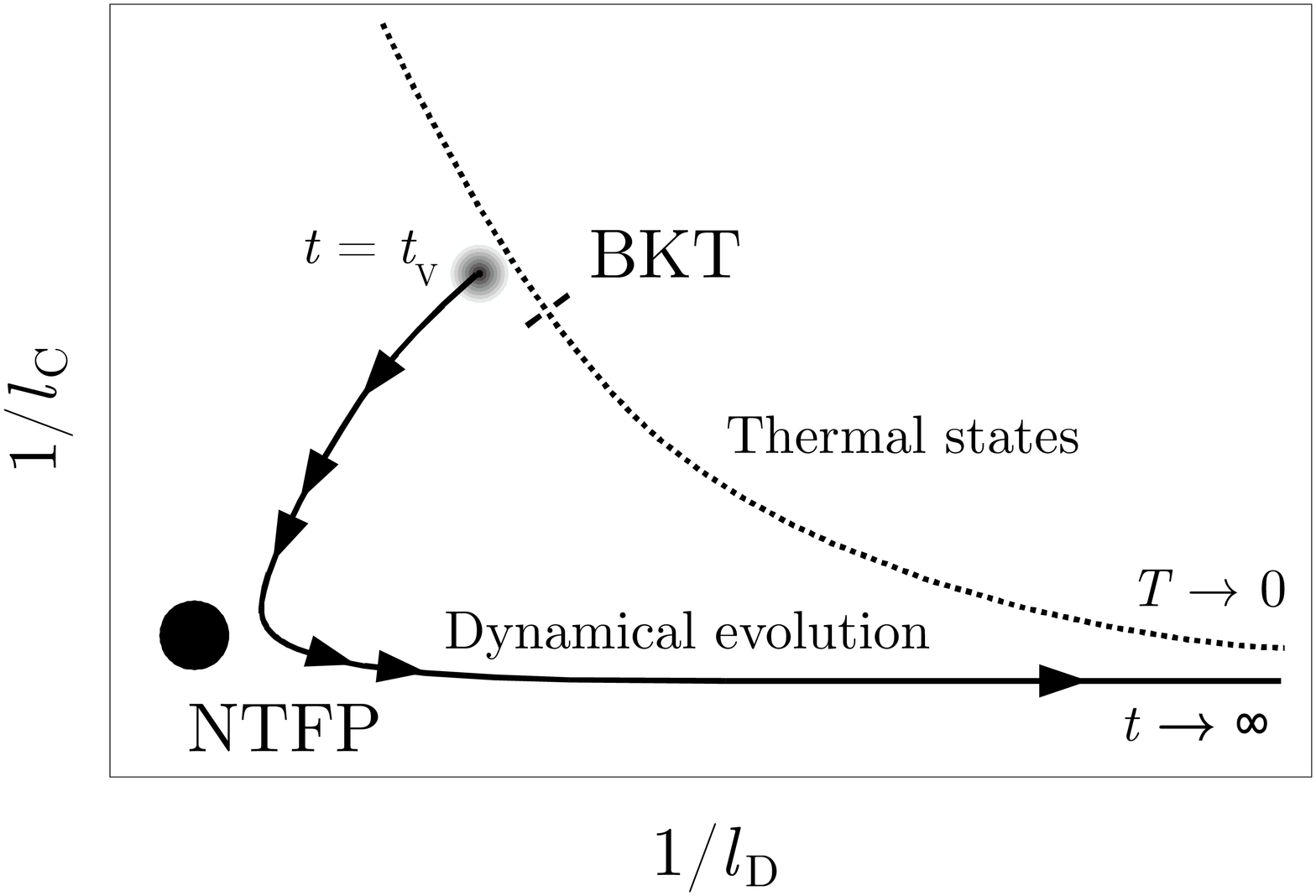}
    \caption{Dynamical evolution of a two-dimensional superfluid near a non-thermal fixed point (NTFP).
     The sketch shows the equilibration process after a quench, in the space of inverse coherence length $l_\mathrm{C}$  and inverse mean vortex-antivortex pair distance $l_\mathrm{D}$. 
    The \textquoteleft Dynamical evolution\textquoteright~ illustrates trajectories of decaying superfluid turbulence starting from the time where vortices appear, ${t}={t}_\mathrm{V}$, approaching the NTFP and finally evolving towards equilibrium.
    The line marked as \textquoteleft Thermal states\textquoteright~qualitatively illustrates these quantities for thermal configurations \cite{Timm1996a, Bisset2009a,Foster2010a}, featuring a steady decrease of inverse coherence with inverse mean vortex-antivortex distance and including a Berezinskii-Kosterlitz-Thouless (BKT) phase transition. 
    An unbinding of vortices of opposite circulation characterises the approach to the NTFP before finally all vortices decay, $l_{\mathrm{D}}\to0$, to establish equilibrium phase coherence, here at a temperature below the BKT transition.}
    \label{fig:BKT2D}
    \end{figure}

In this paper, we study the non-equilibrium dynamics of a two-dimensional Bose gas evolving towards and away from a non-thermal fixed point (NTFP), the stationary properties of which were discussed in Refs.~\cite{Scheppach:2009wu,Nowak:2010tm,Nowak:2011sk}. 
In this process the appearance and decay of vortex excitations play a crucial role. 
We monitor the vortex density during equilibration of the turbulent gas and reveal a bimodal scaling behavior in time. 
By following vortex-antivortex correlations, we show that this phenomenon is directly related to a non-equilibrium vortex unbinding process. Ultimately, vortex excitations evolve into an almost random distribution which constitutes the universal scaling at the NTFP. 

In contrast to classical turbulence, the decay of superfluid turbulence is typically accompanied by a build-up of coherence and quasi-condensation \cite{Levich1978a, Kagan1992a, Kagan1994a, Damle1996a, Berloff2002a, Svistunov2001a, Nazarenko2006a, Kozik2009a}. 
In \Fig{BKT2D}, we sketch the projection of this process onto the space spanned by the coherence length $l_\mathrm{C}$ and the mean inter-vortex pair distance $l_\mathrm{D}$. 
Note that $l_\mathrm{C}$ is defined here in terms of the participation ratio, cf.~\Eq{Coherence}, which is a measure for the width of the normalized first-order coherence function and is less sensitive to noise in the tails of $g^{(1)}(r)$.
In this way, the dynamical evolution towards and away from a non-thermal fixed point can be compared to the properties of near-equilibrium states of a two-dimensional degenerate Bose gas \cite{Onsager1949a,Berezinskii1971a,Kosterlitz1973a, Timm1996a, Bisset2009a, Foster2010a}. 
Arrows mark the direction of the flow and indicate that critical slowing down occurs near the NTFP.
An unbinding of vortices of opposite circulation occurs during the approach of the fixed point before finally all vortices decay to establish full equilibrium phase coherence.

Metastable multi-vortex states or, in one spatial dimension, solitary waves, are also known to appear from strong fluctuations in the vicinity of the normal-fluid to superfluid transition.
Crossing such a transition by varying an equilibrium macroscopic parameter like temperature at a certain rate is well known to induce defect creation. 
Their number depends on the coherence length at the point where the parameter variation ceases to be adiabatic  \cite{Kibble1976a, Zurek1985a}. 
Experiments with ultracold Bose gases following such Kibble-Zurek-type protocols \cite{Weiler2008a,Mathey2012a} as well as generating superfluid turbulence \cite{Henn2009a,Seman2011a} are pursued with increasing effort and could serve to discover and study systematically non-thermal fixed points.

Turbulence has served, since the seminal work of Kolmogorov \cite{Kolmogorov1941a,Obukhov1941a,Frisch1995a}, as one of the first phenomena to develop renormalization-group (RG) techniques out of equilibrium.
The effectively local transport processes in momentum, i.e., scale space, which are at the basis of turbulent cascades immediately suggest themselves for an RG analysis \cite{Eyink1994a}.
For two-dimensional ultracold gases, the dynamical evolution in the vicinity of the Berezinskii-Kosterlitz-Thouless (BKT) critical point \cite{Berezinskii1971a, Kosterlitz1973a} was studied in Refs.~\cite{Mathey2009a},
also in terms of a perturbative RG analysis.
For early work see \cite{Hohenberg1977a} and references cited therein.
A more general set of perturbative non-equilibrium RG equations for the one-dimensional sine-Gordon model near the Luttinger-liquid fixed point was derived in \cite{Mitra2011a}, 
and the route to a non-perturbative analysis also for the strong-coupling regime is provided by out-of-equilibrium functional RG techniques \cite{Canet:2006xu,Gasenzer:2008zz,Gasenzer:2010rq} which will be followed in a forthcoming paper.

%
%
\section{Dynamical simulations}
\label{sec:DynamicalSimulations}
%
%
\begin{figure}[!t]
    \includegraphics[width=0.49\textwidth]{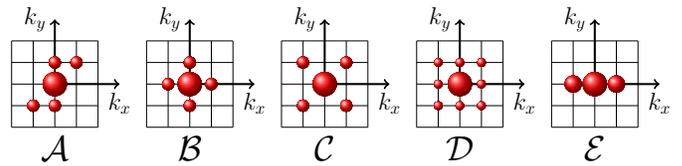}
\caption{(Color online) Panels $\mathcal{A}-\mathcal{E}$ illustrate the initially occupied momentum modes $\vector{k}_\mathrm{ini}$ (shaded discs) with $n(\vector{k}_\mathrm{ini},t=0)\gg1$, for 5 different mean field configurations. The areas of the discs are proportional to the mean number of particles. A sixth initial condition $\mathcal{A}^*$ is geometrically the same as $\mathcal{A}$, but with initially occupied momenta $\vector{k}^*_\mathrm{ini}=4\vector{k}_\mathrm{ini}$. If not stated otherwise, on average half of the total number of particles occupies the zero mode.}
\label{fig:initial2D}
\end{figure}
In this paper we focus on the dynamical evolution of a two-dimensional dilute Bose gas towards a non-thermal fixed point and away from it to thermal equilibrium.
We statistically simulate the far-from-equilibrium dynamics in the classical-wave limit of the underlying quantum field theory. The classical equation of motion for the complex scalar field $\phi(\vector{x},t)$ reads
\begin{equation}  \label{GPE}
   \i \del_t \phi(\mathbf{x},t)= \left[ -\frac{\nabla^2}{2m}+g|\phi(\mathbf{x},t)|^2 \right] \phi(\mathbf{x},t) .
\end{equation}
Here, $m$ is the boson mass, $g$ quantifies the interaction strength in $d=2$ dimensions, and, in our units, $\hbar=1$. Our computations are performed in a computational box of size $L^2$ on a grid with side length $L=N_s a_s$, lattice spacing $a_s$, and periodic boundary conditions. We define the dimensionless variables $\overline{g}=2mg$, $\overline{t}= t/\tau$ with lattice time unit $\tau=2ma^2_s$ and $\overline{\psi}_n(t)=\psi_n a_s \mathrm{exp}(2\i\overline{t})$, see \cite{Nowak:2011sk} for further details. All simulations are performed with parameters $\overline{g}=3\times10^{-5}$ and $N/N^2_\mathrm{s}=1525$,
{where $N$ is the total number of particles}. 
If appropriate, we express length scales in units of the healing length $\xi = (2mgN/L^2)^{-1/2} = 4.6\,a_s$. We drop overbars in the following. 
We choose initial states with a few macroscopically occupied modes in momentum space, as illustrated in \Fig{initial2D}. Fluctuations around these mean values are introduced by sampling the initial field modes according to Gaussian Wigner distributions. Such statistical simulations, being also done under the name truncated Wigner approach, are quasi-exact in the classical wave regime of macroscopic occupation numbers. The type of initial conditions shown  in \Fig{initial2D} induces transport of particles and energy, which leads to vortex creation and turbulence. These field configurations describe quenched superfluid states, which can for instance be prepared by Bragg scattering of photons from a (quasi-)condensate. By varying the number and geometry of initially occupied modes, we can probe the initial-state dependence of our observables. 
%
%
\subsection{Creation of vortices and turbulence}
\label{subsec:MeanFieldDecay}
%
%
%
    \begin{figure}[!t]
   \includegraphics[width=0.47\textwidth]{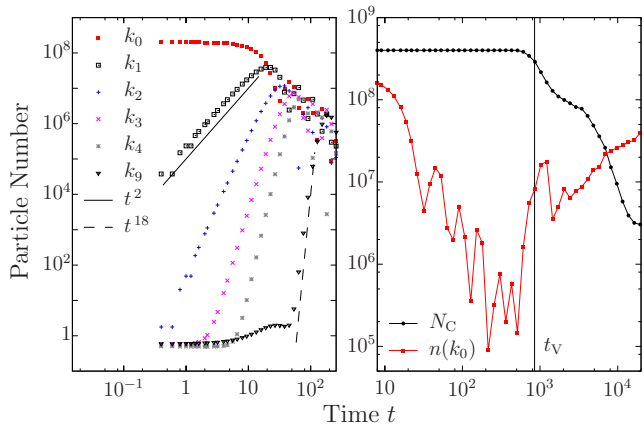}
    \caption{(Color online) Left panel: Single-particle occupation numbers $n(\vector{k},t)=\langle |\phi(\vector{k},t)|^2 \rangle$ as a function of time ${t}$ (in lattice units), for different discrete momentum modes $\mathbf{k}=(k_{n},0)$, $k_n = 2\, \mathrm{sin}(n\pi/N_s)$ along the $k_x$-axis. Results are computed on a grid of size $N_s=512$, from initial field configuration $\mathcal{A}$. Note the double-logarithmic scale. Lines show different power-law evolutions $\sim t^{2n}$. Right panel: Zero-mode occupation number $n((k_0,0),t)$ and coherent population $N_\mathrm{C}= \int\id^2x \, \left|\langle\phi(\vector{x},t)\rangle\right|^2$ as a function of time ${t}$ (double-log.~scale), for an average over $100$ runs, grid size $N_s=512$ and initial condition $\mathcal{A}$. ${t}_\mathrm{V}$ marks the time of vortex creation.}
    \label{fig:ZeroMode2D}
    \end{figure}
%
In the presence of a non-vanishing coupling $g$ the initial states depicted in \Fig{initial2D} are far from thermal equilibrium. During the first stages of the evolution coherent scattering into higher excited modes dominates. In \Fig{ZeroMode2D} (left panel), we show the time evolution of the ensemble-averaged single-particle momentum occupation numbers $n(\vector{k},t)=\langle |\phi(\vector{k},t)|^2 \rangle$ as a function of time, for several momenta $\mathbf{k}=(k_{i},0)$ along the $k_x$-axis. One observes a power-law growth of momenta with $k_x > 0$ until ${t} \simeq 10^2$. This process, exhibiting fast power-law growth $\sim t^{2n}$ of occupations can be understood from analytic mean-field calculations by approximating strong initial occupations to be time-independent. It is present for all our initial conditions and independent of spatial dimension.

In \Fig{ZeroMode2D} (right), we continue to follow the time evolution of the condensate mode, $n(0,t)$. The decay of the zero-mode occupation is part of a nonlocal energy and particle transport to higher momenta. In coordinate space, this process leads to the formation of shock waves, which decay into large numbers of vortices, see Ref.~\cite{Nowak:2011sk} and videos of the evolution~\cite{videos}. In addition, we study the coherent population $N_\mathrm{C}= \int\id^2x \, \left|\langle\phi(\vector{x},t)\rangle\right|^2$ for an initially phase coherent ensemble $N_\mathrm{C}(t=0)=N$. The dynamics preserve coherence until vortices form around time ${t}={t}_\mathrm{V}\simeq 10^3$. This can be understood by considering that the local phase angle $\varphi(\vector{x},t)$ of the complex field $\phi=|\phi|\mathrm{exp}\{\i\varphi\}$ is determined by the positions of the vortices. Since vortices strongly interact, their trajectories in position space quickly randomise. Hence, in the ensemble average the coherent population decays. Beyond this time, density fluctuations are significant only at momenta larger than the inverse healing length, $k>1/\xi$, whereas long-range fluctuations of the Bose gas are dominated by vortical flow. For ${t} \gtrsim {t}_\mathrm{V}$, the zero-mode population $n(0,t)$ starts to increase, signaling the onset of phase ordering associated with vortex annihilations. 
This process will be studied in the following sections.
%
%
\subsection{Vortex density decay}
\label{subsec:VortexDensityDecay}
%
 \begin{figure}[!t]
\flushleft 
(a)\hfill\\[-2ex]  \includegraphics[width=0.45\textwidth]{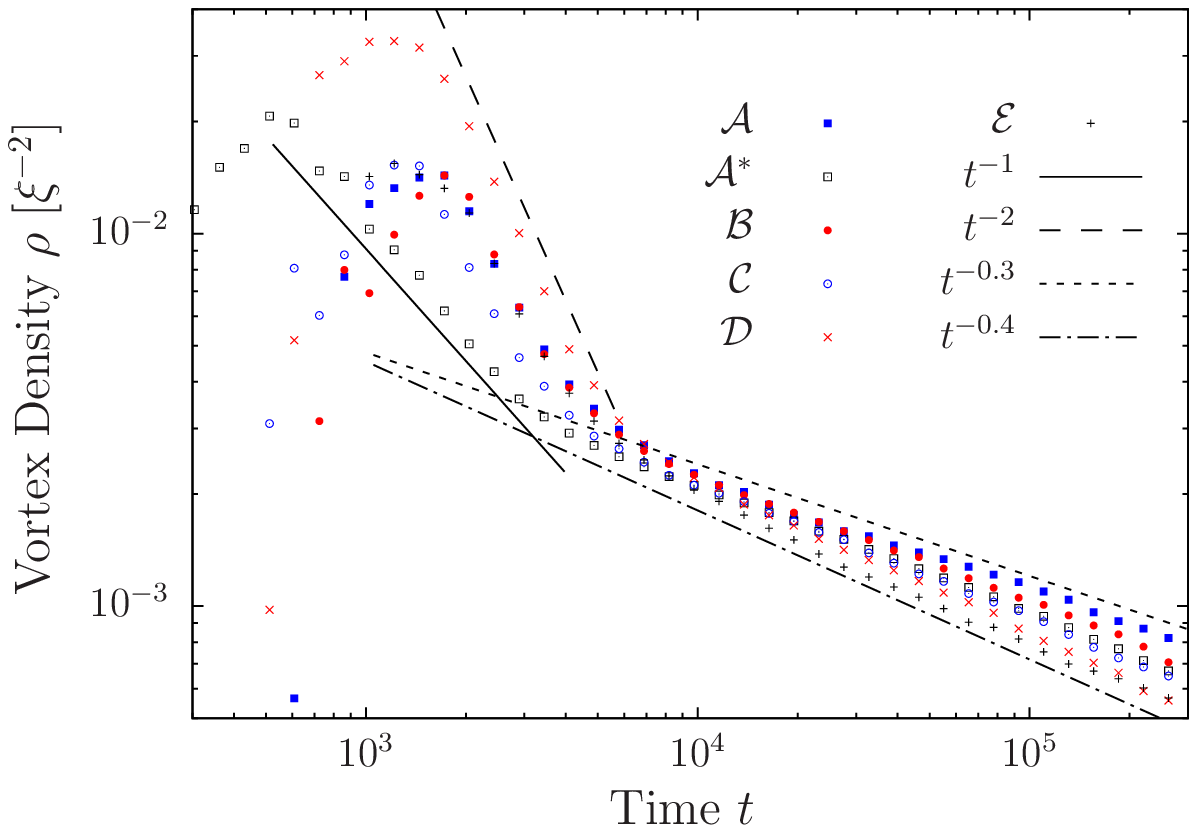}
\\
(b)\hfill\\[-2ex] \includegraphics[width=0.45\textwidth]{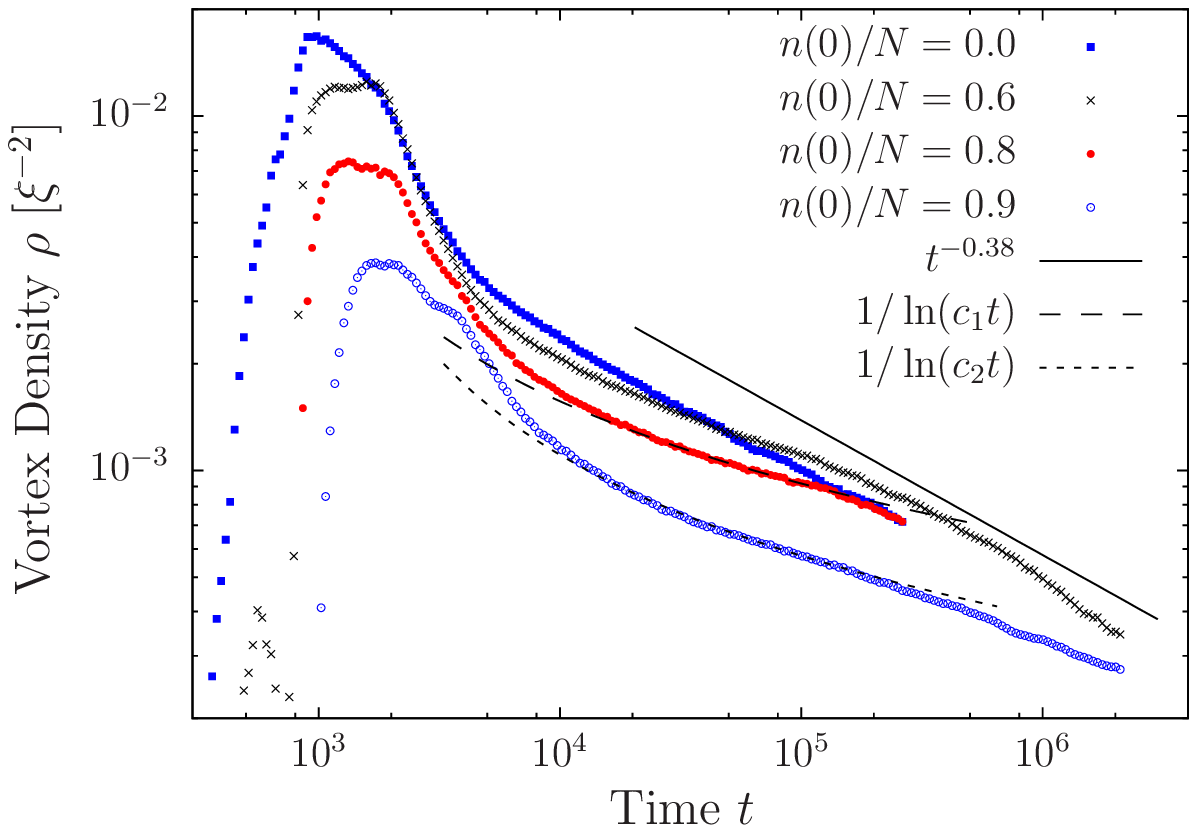}
 \caption{(Color online) Vortex density $\rho$ as a function of time ${t}$ (in lattice units). Note the double logarithmic scale.
 (a) Evolution for various initial conditions as given in \Fig{initial2D}, averaged over 20 runs on a grid of size $N_s=1024$. 
 Lines show different power-law evolutions. Assuming the area of a vortex to be given by $V_\mathrm{V}=(2\xi)^2\pi$, a maximally dense packing of vortices would correspond to the vortex density $\rho_\mathrm{max}=1/V_\mathrm{V}=1/(4\pi\xi^{2})$.
 (b) Evolution for different initial zero-mode populations $n(0)/N$. Averages were taken over 20 runs on a lattice of size  $N_s=1024$, for initial condition $\mathcal{A}$. Fitted parameters are $c_1=0.0026$ and $c_2=0.0012$.
 The closest approach to the NTFP is  reached at $t\simeq(5\dots10)\times 10^{5}$. }
 \label{fig:VortexCount2D2}
 \label{fig:VortexCountImbalance}
 \end{figure}

%
After the creation of vortices, the dynamical evolution exhibits a dual cascade in momentum space, transporting particles from intermediate to small momenta and energy from intermediate to large momenta~\cite{Nowak:2011sk}. The single-particle momentum spectrum develops a quasi-stationary bimodal scaling, with characteristic exponents corresponding to the respective cascade processes. The system approaches a non-thermal fixed point \cite{Scheppach:2009wu, Nowak:2011sk}. The low-momentum scaling of the single-particle momentum distribution can be related to the presence of randomly distributed vortices and antivortices~\cite{Nowak:2011sk}. In the present article, the evolution towards and away from the non-thermal fixed point is investigated. First, we will show that the approach to the non-thermal fixed point is accompanied by a change of the characteristic scaling of the ensemble averaged vortex density $\rho(t)$ with time. 

\Fig{VortexCount2D2}a shows the time evolution of the vortex density
\begin{align}
 \rho(t) 
 &=\langle N^\mathrm{V}(t)+N^\mathrm{A}(t)\rangle/V
\end{align}
with $N^\mathrm{V(A)}(t)$ being the number of vortices (antivortices) in the volume $V$ at time $t$, 
found in simulations starting from the initial conditions defined in \Fig{initial2D}. 
Vortices are counted by detecting their characteristic density and phase profiles. In all runs, vortex formation occurs around ${t}_\mathrm{V}\simeq 10^{3}$, apparent from the steep increase of vortex density around this time. For ${t}\gtrsim {t}_\mathrm{V}$, two distinct stages in the vortex density decay are observed, a rapid early stage and a slow late stage. 
Specifically, the vortex density follows power laws $\rho(t)\sim t^{-\alpha_i}$ with two different exponents $\alpha_i$, $i=1,2$. The exponent during the early stage depends considerably on initial conditions $1 \lesssim \alpha_1 \lesssim 2$, whereas the late stage features a decay exponent in a narrow interval $0.3 \lesssim \alpha_2 \lesssim 0.4$. 
From our analysis given in \Subsect{PhaseCorrelations2D} below, we estimate that the closest approach to the NTFP is  reached at $t\simeq(5\dots10)\times 10^{5}$.

We have repeated our simulations on various grid sizes, $N_s \in \{256, 512, 1024$, $4096\}$. Thereby, we found that decay exponents saturate for and above $N_s=512$. We attribute deviations on smaller grids to effects from regular (integrable) dynamics of few-vortex systems~\cite{Aref1983a}.
We remark that the onset of the slow decay coincides with the development of a particular scaling behavior in the single-particle momentum distribution $n(k) \sim k^{-4}$, which in Refs.~\cite{Scheppach:2009wu, Nowak:2010tm,Nowak:2011sk} was shown to signal the approach to the non-thermal fixed point and the formation of a set of randomly distributed vortices. In this context, the reduction of the vortex density decay exponent, compared to the early stage of rapid decay, is interpreted as a (critical) slowing down of the nonlinear dynamics near the non-thermal fixed point.

As shown in \Fig{VortexCountImbalance}b, the vortex density decay at late times is not always given by a power law. By considerably increasing the initial population of the zero-mode, {e.~g.} $n(0)/N \in \{0.6, 0.8, 0.9\}$, we find that for some time the vortex density is better described by an inverse ln function $\rho(t) \sim 1/\mathrm{ln}(t)$. However, at a late times $t\gtrsim {t}^*$, the decay seems to converge to a power law from the slow-decay regime. By analyzing the dynamics of the single-particle momentum distribution, we could identify the time ${t}^*$ to be the time when compressible excitations have thermalized the high-momentum tail of the spectrum (for details see Fig.~10 in Ref. \cite{Nowak:2011sk}). This is consistent with the observation that the inverse-ln decay could not be observed for initial conditions with small zero-mode occupation where the high-momentum thermalization happens faster.
%
%
\subsection{Vortex correlations}
\label{subsec:VortexCorrelations}
%
%
 \begin{figure}[!t]
\flushleft 
(a)\hfill\\[-2ex] \includegraphics[width=0.45\textwidth]{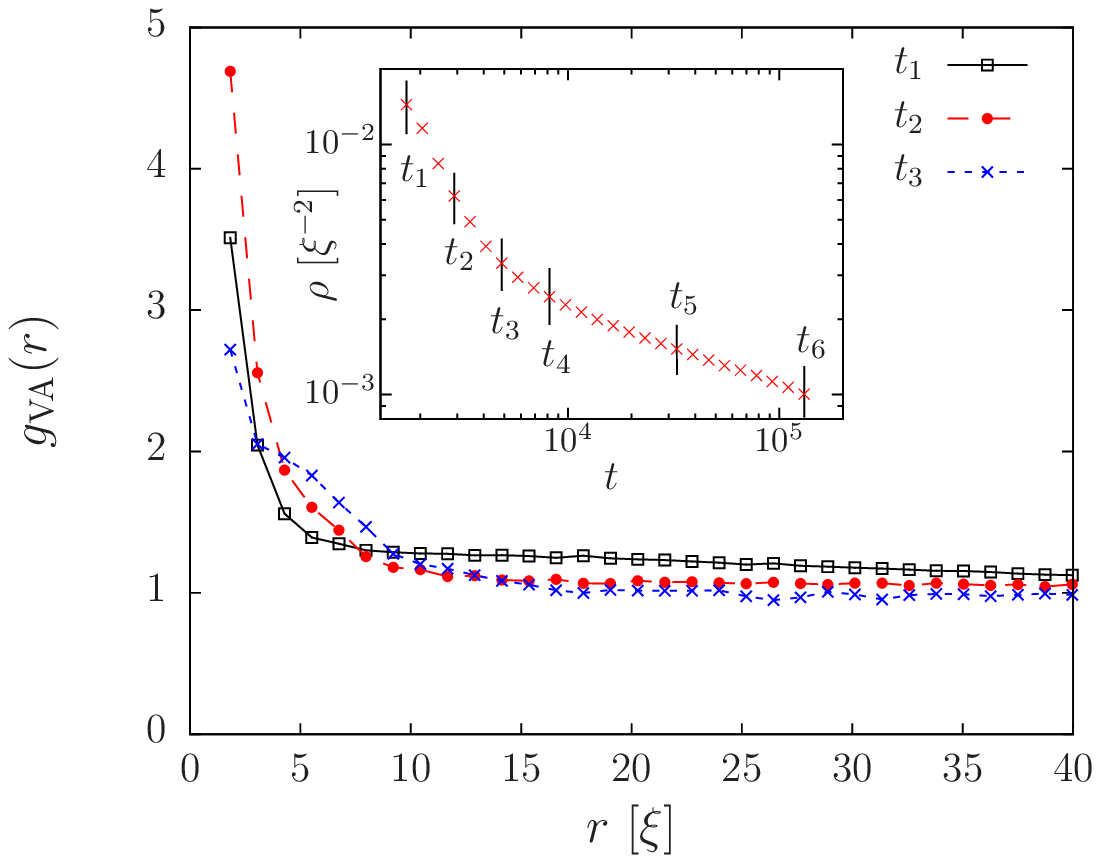}
\\
(b)\hfill\\[-2ex] \includegraphics[width=0.45\textwidth]{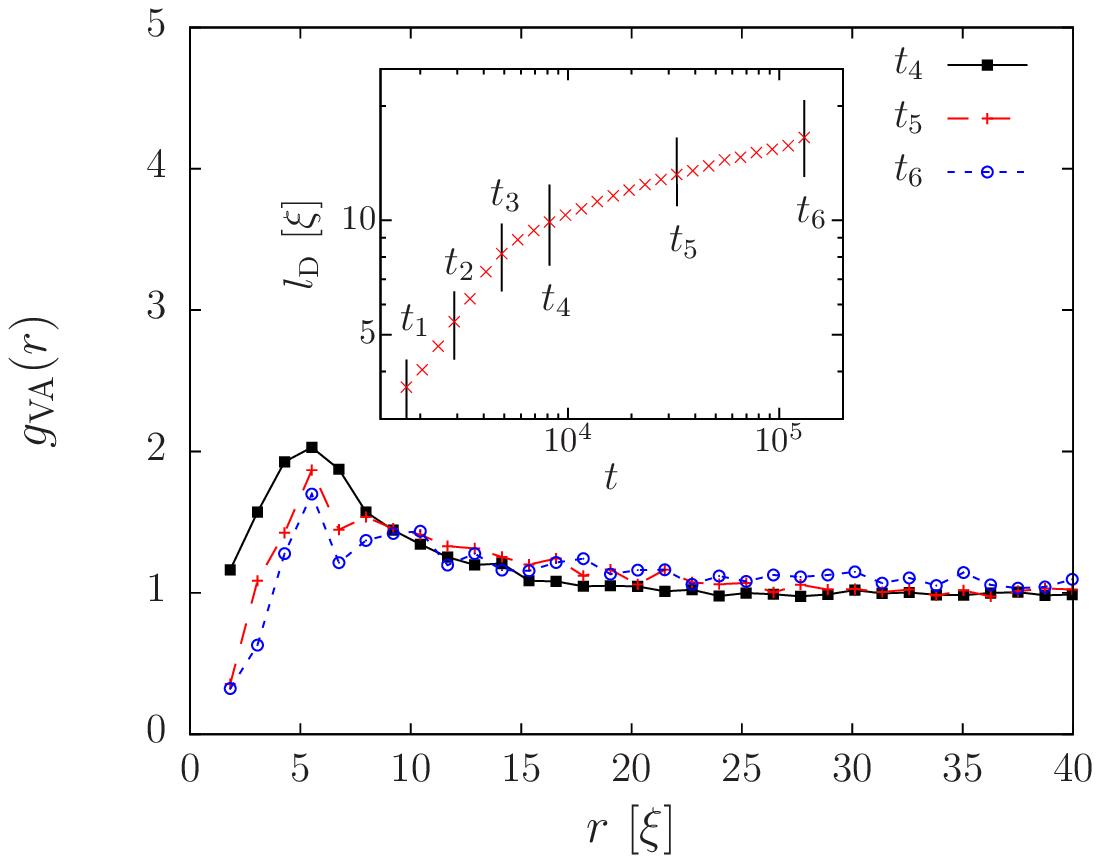}
\caption{(Color online) Normalised vortex-antivortex correlation functions $g_\mathrm{VA}$ defined in \Eq{VortexCorrelation} as a function of radial coordinate $r$, for six different times $t_{i}$ (in lattice units).
(a) $g_\mathrm{VA}(r)$ at times $t_{i}$, $i=1,2,3$ during the rapid-decay stage, averaged over 174 runs on a grid of size $N_s=1024$, using initial condition $\mathcal{A}$. Inset: Vortex density $\rho$ as a function of time $t$, taken from the simulations for \Fig{VortexCount2D2}a.
(b) $g_\mathrm{VA}(r)$ at times $t_{i}$, $i=4,5,6$ during the slow-decay stage, averaged over 174 runs on a grid of size $N_s=1024$, using initial condition $\mathcal{A}$. Inset: Mean vortex-antivortex pair distance $l_\mathrm{D}$ as a function of time, from the simulations for \Fig{VortexCount2D2}a.}
 \label{fig:Vortex_Corr_First_Phase}
 \label{fig:Vortex_Corr_Second_Phase}
 \end{figure}

 \begin{figure}[!t]
 \includegraphics[width=0.45\textwidth]{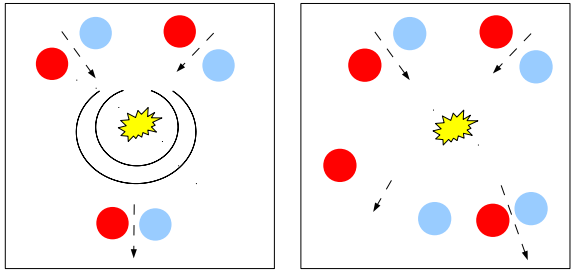}
 \caption{(Color online) Sketch of different scattering events between two vortex-antivortex pairs in $d=2$ dimensions. Left panel: Scattering of two vortex pairs resulting in mutual annihilation of two vortices and the emission of density waves. Right panel: Scattering of two vortex pairs, leading to a change in the vortex anti-vortex distance $l_\mathrm{D}$ and pair velocity $\overline{v}$. We remark, that once $l_\mathrm{D} \sim \xi$ a vortex pair decays rapidly under the emission of density waves.}
 \label{fig:VortexScattering}
 \end{figure}
%
In the following, the dynamical transition in the vortex annihilation dynamics is discussed in terms of characteristic features of the vortex-antivortex correlation function
\begin{equation}
 g_\mathrm{VA}(\vector{x},\vector{x}',t) 
 = \frac{\langle \rho^\mathrm{V}(\vector{x},t)\rho^\mathrm{A}(\vector{x}',t) \rangle}
            {\langle \rho^\mathrm{V}(\vector{x},t) \rangle \langle \rho^\mathrm{A}(\vector{x}',t) \rangle } \, , 
\label{eq:VortexCorrelation}
\end{equation}
where $\rho^\mathrm{V(A)}(\vector{x},t)=\sum_{i}\delta(\mathbf{x}-\mathbf{x}_{i}(t))$ is the distribution of vortices (antivortices) at time $t$ in a single run.
For sufficiently large ensembles, $ g_\mathrm{VA}$ is a function of $r=|\vector{x}-\vector{x'}|$ only.
 
In \Fig{Vortex_Corr_First_Phase}a, we show the evolution of $g_\mathrm{VA}(r,t)$ as a function of $r$ for different times during the fast-decay stage. At early times, one finds a strong pairing peak near $r=0$. 
This peak gets quickly reduced and a hole is \textquoteleft burned\textquoteright~into the correlation function near the origin, see~\Fig{Vortex_Corr_Second_Phase}b. 
Following the time evolution of the spatial vortex distribution we observe that this involves qualitatively different processes:
Mutual annihilations of closely positioned vortices and antivortices occur under the emission of sound waves.
Further separated vortices can approach each other in different ways as illustrated in \Fig{VortexScattering}. 
The scattering of two pairs can directly lead to the annihilation of one pair under the emission of sound waves. 
{We consider this to include events where the pair distance of one dipole reduces below a certain threshold, so that it looks like a density dip rather than a vortex-pair. This density dip can still interact with other vortices, but it will quickly vanish.}
Alternatively, the scattering reduces the vortex-antivortex separation within one pair while it increases it within the other, in accordance with the Onsager point-vortex model \cite{Onsager1949a}.  
We refer to this characteristic change in $g_\mathrm{VA}(r)$ as a vortex unbinding process. 
{The scattering of a closely bound vortex off an isolated vortex is not shown, because it is included as a collision between a closely and a loosely bound pair.}
At around the time ${t}_3 \lesssim {t} \lesssim {t}_4$ the power-law exponent of the vortex density decay changes to about a third of its previous value, see the inset of \Fig{Vortex_Corr_First_Phase}a. 
Next, we compute the mean vortex-antivortex pair distance $l_\mathrm{D}$, by averaging over distances between each vortex and its nearest antivortex. In accordance to the previous discussion, $l_\mathrm{D}$ grows continuously, exhibiting two characteristic stages, see inset of \Fig{Vortex_Corr_Second_Phase}b. At times ${t} \gtrsim 10^4$, $l_\mathrm{D}(t)$ approaches the power-law solution $l_\mathrm{D} \sim \rho^{-1/2}$, as expected for uncorrelated vortices. 
%
%
\subsection{Energy equilibration}
\label{sec:EnergyEquilibration}
%
%
    \begin{figure}[!t]
   \includegraphics[width=0.43\textwidth]{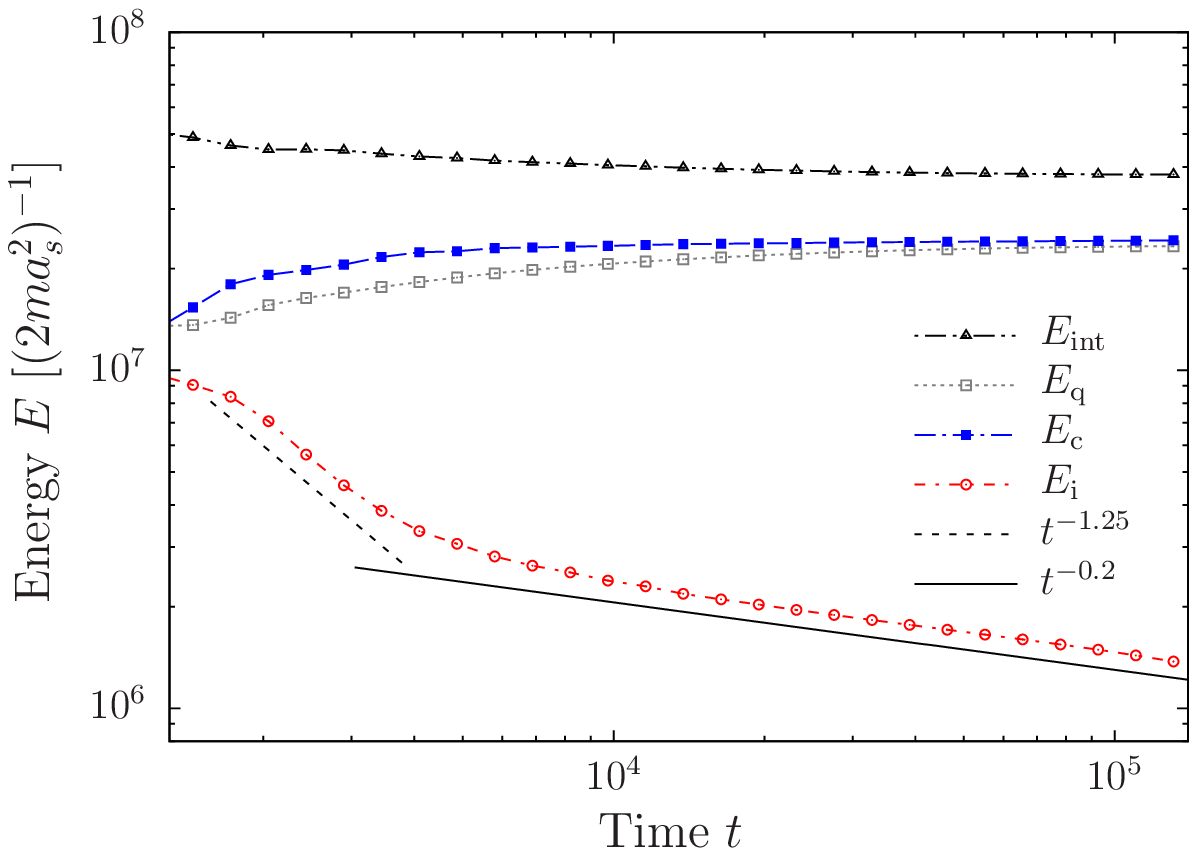}
    \caption{(Color online) Contributions to the total energy as functions of time ${t}$ (in lattice units), averaged over 174 runs. We show the interaction energy $E_\mathrm{int}$, compressible energy $E_\mathrm{c}$, incompressible energy $E_\mathrm{i}$, and quantum pressure $E_\mathrm{q}$, as defined in \App{EnergyDecomposition}, derived on a grid of size $N_s=1024$ and for initial condition $\mathcal{A}$ defined in \Fig{initial2D}. Note the double-logarithmic scale. See \Sect{EnergyEquilibration} for a discussion of the power-law evolutions.}
    \label{fig:Energy2D}
    \end{figure}
%
%
The main result of the previous section was the relation of different stages in the evolution of the vortex density $\rho(t)$ to characteristic features in the vortex-antivortex correlation function  $g_\mathrm{VA}(r,t)$. The vortex density decay, was shown to be accompanied by a dynamical vortex unbinding. This finding can be supplemented by considering the evolution of different energies contained in the gas. In Refs.~\cite{Nore1997a,Nore1997b} it was suggested to decompose the kinetic energy into an incompressible and a compressible component, which, for conciseness, we give details of in \App{EnergyDecomposition}. In this way, contributions from vortical excitations can be separated from other excitations such as sound waves. In \Fig{Energy2D}, we show the time evolution of different energy components for initial condition $\mathcal{A}$. One observes that the decay of the incompressible energy during the early-time stage {can be estimated to follow a power-law} $\sim t^{-1.25}$ and in the late-time stage $\sim t^{-0.2}$. This decay happens considerably slower than the vortex density decay in the early- (late-)time stage $\sim t^{-1.7} (\sim t^{-0.3})$, discussed in Sect.~\subsect{VortexDensityDecay}. As a result, the incompressible energy per vortex grows as $\sim t^{0.45}$ at early times and as $\sim t^{0.1}$ at late times. Since the energy of a vortex pair increases with distance, this is in agreement with the phenomenon of increasing vortex-antivortex pair distance $l_\mathrm{D}$. 

At late times, compressible and quantum-pressure energy components develop into an equipartitioned state, also observed in decaying superfluid turbulence starting from a Taylor-Green vortex configuration \cite{Krstulovic2011a}.
%

\subsection{Possible relation to classical turbulence \\ in $d=2$ dimensions}
\label{subsec:ClassTurbulence}
In the following, we discuss a similarity between our results and findings in classical fluid turbulence. Great efforts have been made to investigate freely decaying turbulence in two-dimensional classical fluids, see, {e.g.}, Refs.~\cite{Carnevale1991a,Weiss1993a,Bracco2000a,Tabeling:decay:1991,Benzi:decay:1992,Sire:merging:2010,Sire:threebody:2000,Yakhot:crossover:2004}. In this context, special focus was set on the decay of the enstrophy $\Omega(t)$, which is related to the vorticity $\vector{\omega}=\nabla \times  \vector{v}$ of the velocity field $\vector{v}(\vector{x},t)$ by
\begin{equation}
\Omega(t)=\frac{1}{2} \int \mathrm{d}^2x \, |\vector{\omega}(\vector{x},t)|^2 \,.
\end{equation}
It was found by different methods that the long-time decay of the enstrophy is given by a power-law $\Omega(t)\sim t^{-\gamma}$, with $\gamma \simeq 0.35-0.4$~\cite{Carnevale1991a,Weiss1993a,Bracco2000a,Tabeling:decay:1991,Benzi:decay:1992,Sire:merging:2010,Sire:threebody:2000,Yakhot:crossover:2004}.

In superfluids, vorticity is concentrated in the vortex cores. The vorticity of a turbulent flow consisting of $M$ vortices with circulations $\kappa_i$ and positions $\vector{x}_i$, for $i < M$, is given by $\vector{\omega}(\vector{x},t) = \sum_i \kappa_i \delta( \vector{x}-\vector{x}_i(t) ) $. Hence, in a flow consisting of vortices with circulations $\kappa_i=\pm 1$ the enstrophy reads
\begin{equation}
\Omega(t)=\frac{1}{2}\delta(\vector{0})M(t)\,,
\end{equation}
which is, with $\delta(\vector{0})\sim 1/V$, proportional to the vortex density,  $\Omega(t) \varpropto \rho(t)$. As shown in \Fig{VortexCount2D2}a, in the late-time stage, our results are in accordance with the results from classical turbulence. However, we point out that the mechanisms of enstrophy decay in the two systems are fundamentally different. Whereas in superfluids vortices annihilate, the main process of vorticity decay in classical two-dimensional fluids is the merging into larger vortices. 

We finally remark that in numerical simulations of freely decaying classical turbulence a crossover between two stages of power-law decay similar to our findings has been reported in Ref.~\cite{Yakhot:crossover:2004}.

\subsection{Kinetic theory of vortex scattering}
\label{subsec:KineticTheory}
%
%
 \begin{figure}[!t]
 \includegraphics[width=0.48\textwidth]{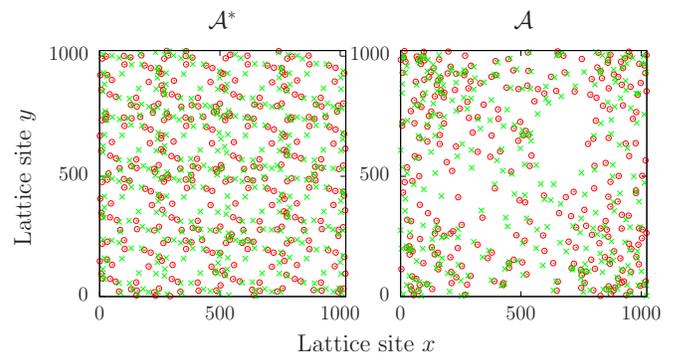}
 \caption{(Color online) Vortex and antivortex positions shortly after turbulence creation at ${t}\simeq{t}_\mathrm{V}$, for two single runs of the simulations on a grid of size $N_s=1024$. Left panel: Initial condition $\mathcal{A^*}$. Right panel: Initial condition $\mathcal{A}$, as defined in \Fig{initial2D}.}
 \label{fig:VortexPositions}
 \end{figure}
%
The decay of the vortex density has been investigated in two-dimensional classical fluids \cite{Carnevale1991a,Weiss1993a,Bracco2000a,Tabeling:decay:1991,Benzi:decay:1992,Sire:threebody:2000,Sire:merging:2010,Yakhot:crossover:2004} and superfluids~\cite{Chu2001a,Nazarenko2006a,Nazarenko2007a, Numasato2010a}, mainly in the presence of driving or dissipation. Several authors have proposed kinetic theories building upon assumptions about the decay process \cite{Ambegaokar1980a, Carnevale1991a, Sire:threebody:2000, Sire:merging:2010}.
{The vortex decay cannot be explained by a simple model of independent vortices and antivortices moving towards each other to minimize the energy. Neglecting interactions with sound  vortex-antivortex pairs perform a collective motion perpendicular to their relative distance vector without changing their distance.}
This motion quickly leads to pair-pair collisions and establishes a kinetic-theory picture for vortex pairs. In \Fig{VortexScattering}, we showed two examples of vortex-pair scattering processes altering the vortex density $\rho(t)$ and vortex correlation functions $g_\mathrm{VA}(r,t)$.

Assuming that the vortices are moving in pairs and that annihilations happen as the result of collisions of vortex pairs, the decay rate for the number $N_\mathrm{V}$ of vortices follows from the number of dipoles $N_\mathrm{D}\sim N_\mathrm{V}$ and $\partial_t N_\mathrm{D}(t) \sim - N_\mathrm{D}(t)/\tau$, with mean free collision time $\tau$. The mean free collison time $\tau = l_\mathrm{mfp}/\bar v$ is given by the mean velocity of the pairs $\bar v$ and the mean free path $l_\mathrm{mfp}=V/(\sigma N_\mathrm{D})$ with cross section $\sigma$. Both, mean velocity and cross section depend on the number of vortices via the mean vortex-antivortex pair distance $l_\mathrm{D}$ according to $\bar v \sim 1/l_\mathrm{D}$ and $\sigma \sim l_\mathrm{D}$. These considerations result in a rate equation
\begin{equation}
  \partial_t N_\mathrm{V}(t) = - c N_\mathrm{V}^2 /\tau
\label{eq:KinTheory}
\end{equation}
with dimensionless constant $c$. \Eq{KinTheory} has the power-law solution $N_\mathrm{V}(t) \sim t^{-1}$. This decay law can only be observed for specific initial conditions $\mathcal{A^*}$ in the early-time stage. We attribute deviations from $t^{-1}$ scaling during this period to an inhomogeneous distribution of vortices, encountered for certain initial conditions. To give an example, we show the vortex distributions created from the two different initial conditions $\mathcal{A^*}$ and $\mathcal{A}$ in \Fig{VortexPositions}. 

During the late-time stage, only a few vortices are bound in pairs, while most vortices are loosely bound and interact equally with a larger number of vortices around them. We heuristically take this into account by considering a modified scattering cross section $\sigma \sim l_\mathrm{D}N^2_\mathrm{D}$. The resulting kinetic equation reads $\partial_t N_\mathrm{V}(t) \sim - N_\mathrm{V}^4$, with solution $N_\mathrm{V}(t) \sim t^{-1/3}$, and hence yields the reduction of the decay exponent observed at late times in our simulations.
%
%
%
\subsection{Phase correlations}
\label{subsec:PhaseCorrelations2D}
%
    \begin{figure}[!t]
   \includegraphics[width=0.45\textwidth]{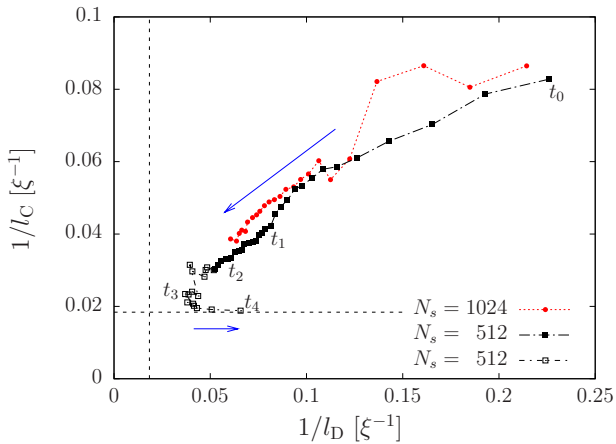}
    \caption{(Color online) Trajectories of multi-vortex states in the space of inverse coherence length $1/l_\mathrm{C}$ and inverse mean vortex-antivortex distance $1/l_\mathrm{D}$, starting from $t=t_\mathrm{V}$. Arrows are added to guide the eye along the time direction. Dashed lines mark the minimal values $2/L=0.018\xi^{-1}$, available on a grid of size $N_s=512$. 
    Our understanding of the NTFP as a configuration with a few, maximally separated pairs on an otherwise maximally coherent background implies it to be located near the crossing of the dashed lines.
    Hence, the NTFP is approached most closely between $t\simeq 5\times 10^{5}$ and $t\simeq10^{6}=t_{3}$.
    Closed (red) circles mark an average over 174 runs, $N_s=1024$, initial time ${t}_\mathrm{V}=2.3\times10^{3}$ and final time ${t}_\mathrm{f}= 1.3\times10^{5}$; 
    closed squares an average over 1223 runs, with $N_s=512$, ${t}_\mathrm{V}=1.7\times10^{3}$ and ${t}_\mathrm{f}= 2.6\times10^{5}$; open squares an average over 16 runs, $N_s=512$, ${t}_\mathrm{i}=2.6\times10^{5}$, and ${t}_\mathrm{f}= 4.2\times10^{6}$. 
    Note that the symbols are equally spaced on a logarithmic time scale. 
    We indicate the times $t_0 = 1.7\times10^{3},$ $t_1 = 1.6\times10^{4}$, $t_2 = 1.3\times10^{5}$, $t_3 
= 1.0\times10^{6}$, $t_4 = 4.2\times10^{6}$. 
{The average number $N_\mathrm{V}(t)$ of vortices left in the system is $N_\mathrm{V}(t_0)=97.7$, $N_\mathrm{V}(t_1)=21.8$, $N_\mathrm{V}(t_2)=11.7$, $N_\mathrm{V}(t_3)=5.8$ and $N_\mathrm{V}(t_4)=2.1$.}}
    \label{fig:Correlations2D}
    \end{figure}
%
    \begin{figure}[!t]
 \flushleft 
(a)\hfill\\[-2ex] \includegraphics[width=0.45\textwidth]{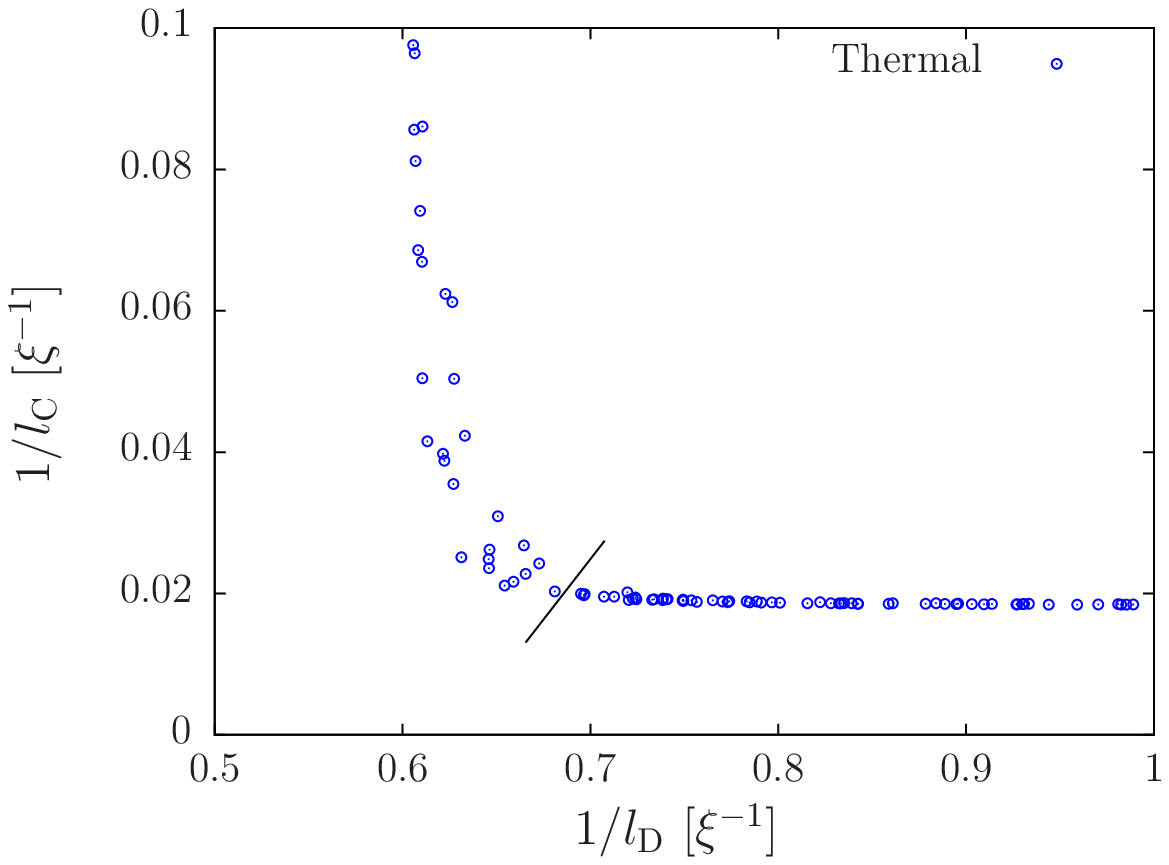}
\\
(b)\hfill\\[-2ex] \includegraphics[width=0.45\textwidth]{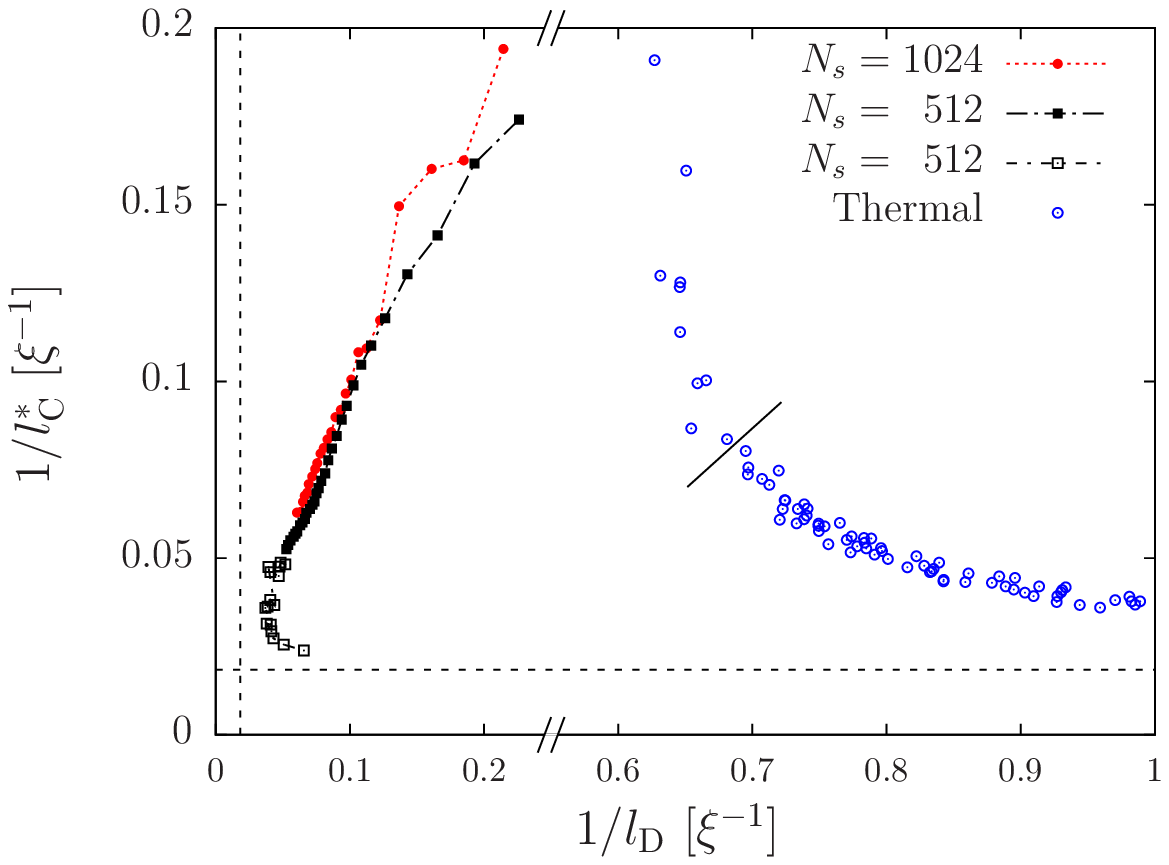}
\caption{{(Color online) Multi-vortex states in the space of inverse coherence length and inverse mean vortex-antivortex distance. 
(a) Thermal configurations $(l_\mathrm{D}^{-1}(T), l_\mathrm{C}^{-1}(T))$ for a range of temperatures $T$, increasing from bottom right to top left. The solid line marks the point where the decay of the $g^{(1)}(r)$-function changes from algebraic to exponential, signaling the BKT-transition. 
(b) Comparison of the thermal line $(l_\mathrm{D}^{-1}(T), l_\mathrm{C}^{*-1}(T))$ for the same range of temperatures $T$ with the corresponding dynamical evolution. (Same data as in \Fig{Correlations2D}.) Dashed lines mark the minimal values $2/L=0.018\xi^{-1}$, available on a grid of size $N_s=512$. Note that the $(1/l_{\mathrm{D}})$-axis interval [0.25,0.55] has been cut out.}}
    \label{fig:NTFPvsThermal}
    \end{figure}
%
    \begin{figure}[!t]
 \includegraphics[width=0.42\textwidth]{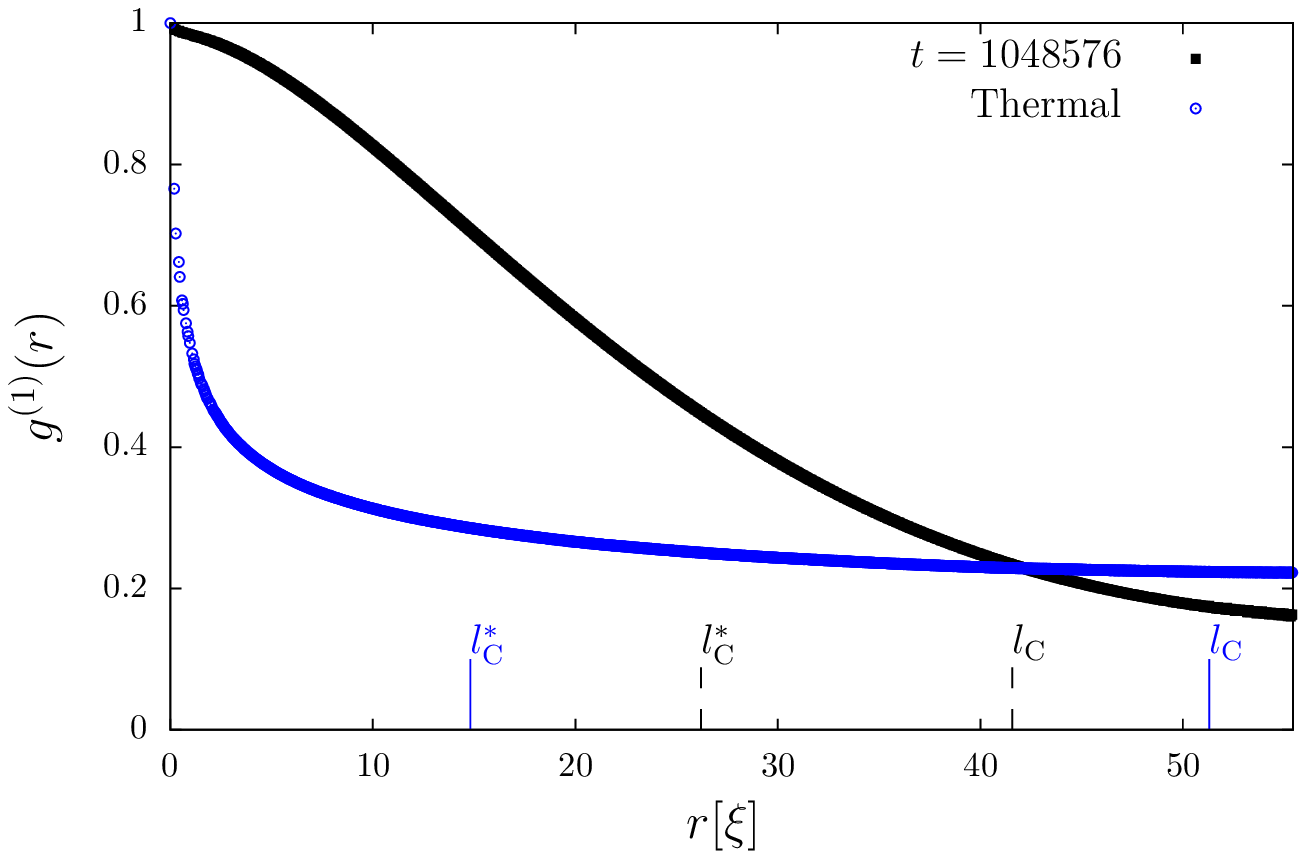}
\caption{(Color online) First order coherence function $g^{(1)}$ as a function of radial coordinate $r$. We show one example for the system during the equilibration process at time $t=1.05\times10^{6}$, with $l_{\mathrm{C}}=41.6\xi$, $l^{*}_{\mathrm{C}}=26.2\xi$, and one for a thermal configuration, giving $l_{\mathrm{C}}=51.3\xi$, $l^{*}_{\mathrm{C}}=14.8\xi$. 
}
    \label{fig:g1functions}
    \end{figure}
%
In the remainder of this article, we focus on the growth of long-range coherence at late times, associated with the annihilation of topological defects \cite{Levich1978a, Kagan1992a, Kagan1994a, Damle1996a, Berloff2002a, Svistunov2001a, Kozik2009a, Nazarenko2006a}. From this point of view freely decaying superfluid turbulence is a particular example of phase-ordering dynamics after a quench into the ordered phase \cite{Bray1994a}. Whereas in three dimensions a second-order phase transition connects a normal-fluid and a superfluid phase, a Bose gas in two dimensions experiences a Berezinskii-Kosterlitz-Thouless (BKT) transition \cite{Berezinskii1971a, Kosterlitz1973a}. For the two-dimensional ultracold Bose gas, experimental and theoretical results support the understanding of the phase transition in terms of vortices undergoing an unbinding-binding transition \cite{Hadzibabic2006a, Simula2006a, Schweikhard2007a, Giorgetti2007a, Weiler2008a, Bisset2009a, Foster2010a}. 

In this context, we are interested in a comparison between correlation properties observed in the non-equilibrium dynamics near a non-thermal fixed point and those known from equilibrium studies. We compute the dynamical trajectory of the vortex gas in the space of inverse coherence length and inverse mean vortex-antivortex pair distance. 
{We compare our results to simulations of a thermal two-dimensional Bose gas specifically for our system parameters. For this, we evolve field configurations in time which are initially close to a thermal Rayleigh-Jeans distribution at a temperature $T$. After equilibration is reached, we compute the position of the states in the above phase space for different $T$.}

We define a coherence length $\l_\mathrm{C}$ in terms of the participation ratio \cite{Kramer1993a} of the angle-averaged first-order coherence function $g^{(1)}(r) = \int \mathrm{d}\theta \, \langle \phi^*(\vector{x})\phi(\vector{x}+\vector{r}) \rangle /\sqrt{\langle n(\mathbf{x})\rangle\langle n(\mathbf{x}+\mathbf{r})\rangle} $,
\begin{equation}
\l_\mathrm{C} = \left( \mathcal{N}\int \mathrm{d}r \, [g^{(1)}(r)]^2 \right)^{-1} \,, 
\label{eq:Coherence}
\end{equation}
with $\mathcal{N}=[\int \mathrm{d}r g^{(1)}(r)]^{-2}$.
It measures the spatial extension of the first order coherence function. Other than $r_\mathrm{coh} = \int \mathrm{d}{r} \, r^{2} g^{(1)}(r)/\int \mathrm{d}{r}\, r\,g^{(1)}(r)$, the quantity $\l_\mathrm{C}$ does not sum up values of $g^{(1)}(r)$ weighted by the distance, which would enlarge insignificant contributions at large $r$. In addition, it gives meaningful results also in the case of large coherence $ g^{(1)}(r) \simeq 1$, where for instance the FWHM-measure can not be applied any more. Note that in equilibrium this quantity smoothly interpolates between the regime of exponential decay of $g^{(1)}$ above the BKT-transition, where the exponential coherence length $\xi_\mathrm{C}$ is defined as $ g^{(1)}(r) \sim \mathrm{exp}( -r/\xi_\mathrm{C} )$, and its power-law decay in the superfluid regime. 

In \Fig{Correlations2D}, we follow the time evolution of the gas for ${t} >{t}_\mathrm{V}$. One can observe, that a state of low coherence and small mean vortex-antivortex pair distance evolves towards larger coherence and larger vortex-antivortex separation. As discussed in Sects.~\ref{subsec:VortexDensityDecay} and~\ref{subsec:VortexCorrelations}, this is due to vortex annihilations and vortex-antivortex unbinding. For times ${t} > 10^4$, the coherence length grows as $l_\mathrm{C} \sim \rho^{-1/2}$, in the same way as $l_\mathrm{D}$ shown in \Fig{Vortex_Corr_Second_Phase}b. The evolution considerably slows down for $1/l_\mathrm{C} \sim 1/l_\mathrm{D} \rightarrow 0$. In this regime, the Bose gas shows characteristic scaling properties, see Refs.~\cite{Nowak:2010tm,Nowak:2011sk}, which indicate the presence of the non-thermal fixed point~\cite{Scheppach:2009wu}. After spending a long time near this point, the mean vortex-antivortex pair distance declines. This is a sign that the last remaining vortex-antivortex pairs reduce their size prior to their annihilation and the equilibration of the system.
{Around the same time the power-law in the vortex density decay shown in \Fig{VortexCount2D2} breaks down.}

    Our understanding of the NTFP as a configuration with a few, maximally separated pairs on an otherwise maximally coherent background implies it to be located near the crossing of the dashed lines.
    Hence, the NTFP is approached most closely between $t\simeq 5\times 10^{5}$ and $t\simeq10^{6}=t_{3}$.

{To set the above evolution in relation to equilibrium configurations, we show, in \Fig{NTFPvsThermal}a, the thermal line $(l_\mathrm{D}^{-1}(T), l_\mathrm{C}^{-1}(T))$ for a range of temperatures $T$ for which the system shows a non-vanishing zero-mode population.
Note that, in order to define what counts as a bound vortex pair, we filter out field fluctuations on scales smaller than $0.55\xi$ before detecting vortices and antivortices. Hence, the resulting inverse of the mean vortex-antivortex pair distance represents a lower bound, and the  separation of the NTFP from the thermal configurations becomes obvious.

In view of the thermal results it is useful to consider an alternative definition for the coherence length. The length $\l^*_\mathrm{C}= \int \mathrm{d}r \,g^{(1)}(r)$ shares the above mentioned advantages of the participation ratio. In addition, it does not overestimate the coherence of flat distributions. The thermal $g^{(1)}(r)$ functions show a fast decay at short distances $r_s$ to a value $g^{(1)}_c$ and are almost constant for $r>r_s$. 
\Fig{g1functions} shows two typical $g^{(1)}$-functions, one for the system during the equilibration process at time $t=1.05\times10^{6}$,  with $l_{\mathrm{C}}=41.6\xi$, $l^{*}_{\mathrm{C}}=26.2\xi$, and one for a thermal configuration, giving $l_{\mathrm{C}}=51.3\xi$, $l^{*}_{\mathrm{C}}=14.8\xi$.
Due to the small area under $g^{(1)}(r)$ up to $r_s$ the normalization $\mathcal{N}$ in the definition of the participation ratio enlarges the integrand of \Eq{Coherence} close to unity. Hence, $l_\mathrm{C}$ becomes large while $\l^*_\mathrm{C}$ is less affected by this effect. 
In \Fig{NTFPvsThermal}b we compare the thermal line $(l_\mathrm{D}^{-1}(T), l_\mathrm{C}^{*-1}(T))$ for the same range of temperatures $T$ with the corresponding dynamical evolution.
}

In \Fig{BKT2D}, our findings are summarized qualitatively in a reduced phase space of the vortex gas. 
In this way, the \textquoteleft Dynamical evolution\textquoteright~of decaying superfluid turbulence is compared to an estimate of the expected equilibrium configurations illustrated by the dashed line marked \textquoteleft Thermal states\textquoteright~along which the temperature $T$ varies. The most significant difference between these two lines is that the non-equilibrium dynamics is characterised by an increase of the mean vortex-antivortex pair distance with increasing coherence, whereas equilibrium configurations are expected to feature a decrease in pair distance with increasing coherence. A slow-down of the dynamics together with characteristic scaling of the single-particle momentum distribution, observed in the regime of large coherence and large vortex-antivortex pair distance, marks the position of the non-thermal fixed point. Finally, when all vortices have annihilated, the system reaches the \textquoteleft Thermal states\textquoteright-line deep in the superfluid regime.

\section{Conclusions}
\label{sec:summary}
We have studied the non-equilibrium dynamics of a two-dimensional dilute ultracold Bose gas towards and away from a non-thermal fixed point  (NTFP). 
Following an initial quench, evolution towards a fixed point appears to be a generic feature of the (quasi-)conden\-sa\-tion process and the build-up of coherence.
In the course of a critically slowed evolution, vortex excitations evolved into an almost random distribution reflected in the scaling of the single-particle spectra at the NTFP.
We showed that the vortex-density decay is directly related to a non-equilibrium vortex unbinding process. 

Our results allow to draw a picture of the NTFP.
The evolution path towards and away from the fixed point is summarized schematically in \Fig{BKT2D}. 
There, we sketch the evolution of the multi-vortex states in the plane spanned by the coherence length $l_\mathrm{C}$, defined in terms of the participation ratio, and the mean inter-vortex pair distance $l_\mathrm{D}$.
This picture allows to compare our results with quasi-equilibrium studies of a two-dimensional vortex gas. 
The NTFP emerges to bear similarities to the equilibrium Berezinskii-Kosterlitz-Thouless (BKT) fixed point.
The NTFP is characterized by a few, in the extreme case, one pair of far separated anti-circulating vortices.
While the BKT transition also features unbinding of vortices, finite temperature implies the simultaneous excitation of many rotons, i.e., strongly bound vortex-antivortex pairs.
The NTFP is identified by strong wave turbulent scaling in the infrared limit \cite{Scheppach:2009wu}, $n(k)\sim k^{-4}$.
At the same time, the high-energy modes can be populated in a much weaker way, e.g., at considerably lower temperature than the BKT critical temperature or remain out of equilibrium.
The details of the UV mode populations are determined by the way how the NTFP is being approached.
We emphasize that the approach of the NTFP is a generic but out-of-equilibrium process and that eventually, the system will decay to an equilibrium state potentially far away from the NTFP.
In the examples discussed in this paper, the total energy is sufficiently low such that the final equilibrium state emerges to be considerably below the BKT critical temperature.

The way we force the system here as in the work reported in \cite{Nowak:2010tm,Nowak:2011sk} to  approach the NTFP generalizes that of Kibble and Zurek. 
A strong sudden quench replaces the more or less adiabatic approach of the BKT phase transition. 
To what extent the BKT phase transition can be understood as happening within a class of thermal states near the NTFP studied here needs to be clarified by analyzing full out-of-equilibrium renormalisation-group equations in comparison with standard descriptions in thermal equilibrium.
We point out that the concept of an NTFP is far more general than the specific situation studied here, see, e.g., Refs.~\cite{Berges:2008wm,Berges:2008sr,Berges:2008mr,Scheppach:2009wu,Berges:2010ez,Gasenzer:2011by,Carrington:2010sz,Schmidt:2012kw}.

Further important questions concern the relation to fully developed quantum turbulence \cite{Nore1997a,Vinen2002a,Araki2002a,Kobayashi2005a,Proment2010a,Henn2009a,Seman2011a,Neely2012a} which is believed to exhibit a quasi-classical Kolmogorov-Obukhov scaling \cite{Kolmogorov1941a,Obukhov1941a} of the radial energy, $E(k)\sim k^{-5/3}$, in the infrared regime below the mean inverse distance between vortices \cite{Nore1997a,Vinen2002a,Araki2002a,Kobayashi2005a,Proment2010a}. 

In a recent experiment, a strong reduction of the vortex decay rate at late times has been reported \cite{Neely2012a}.
Quantitative experimental observation of the predictions made here could give new insight into the character of the NTFP and its relation to quantum turbulence.

%
\acknowledgments
The authors thank B.~Anderson, J.~Berges, N.~G.~Ber\-loff, M.~J.~Davis, G.~Krstulovic, J.~M. Pawlowski, D.~Sexty, B.~Svistunov, and M.~Tsubota for useful discussions. 
They acknowledge support by the Deut\-sche Forschungsgemeinschaft (GA677/7,8), by the University of Heidelberg (Excellence Initiative, Center for Quantum Dynamics), by the Helmholtz Association (HA216/EMMI), and by the University of Leipzig (Grawp-Cluster).

\appendix
\label{sec:appendix}
%
\section{Quantum hydrodynamic energy decomposition}
\label{app:EnergyDecomposition}
In this appendix we give details of the energy decomposition introduced in Refs.~\cite{Nore1997a,Nore1997b}.
To exhibit vortical flow and define the decomposition we use the polar representation $\phi(\mathbf{x},t)=\sqrt{n(\mathbf{x},t)}\exp\{i\varphi(\mathbf{x},t)\}$ of the field in terms of the density $n(\mathbf{x},t)$ and a phase angle $\varphi(\mathbf{x},t)$. 
This allows to express the particle current $\mathbf{j}=i(\phi^*\nabla \phi - \phi \nabla \phi^*)/2=n\mathbf{v}$ in terms of the velocity field $\mathbf{v}=\nabla\varphi$.
With this, we decompose the kinetic-energy spectrum following Ref.~\cite{Nore1997a,Nore1997b}, splitting the total kinetic energy $E_{\mathrm{kin}}= \int \mathrm{d}^dx \, \langle |\nabla \phi(\mathbf{x},t)|^2\rangle/(2m)$ as $E_{\mathrm{kin}} = E_{\mathrm{v}} + E_\mathrm{q}$ into a `classical' part and a `quantum-pressure' component, 
\begin{align}
\label{eq:Ev}
E_\mathrm{v}
&= \frac{1}{2m}\int \mathrm{d}^dx \, \langle |\sqrt{n}\mathbf{v}|^2 \rangle ,
\\
E_\mathrm{q}
&=\frac{1}{2m}\int \mathrm{d}^dx \, \langle |\nabla \sqrt{n}|^2 \rangle \,.
\end{align}
The radial energy spectra for these fractions involve the Fourier transform of the generalised velocities $\vector{w}_{\mathrm{v}}=\sqrt{n}\vector{v}$ and $\vector{w}_{\mathrm{q}}=\nabla\sqrt{n}$,
\begin{eqnarray}
 E_{\delta}(k)= \frac{1}{2m} \int k^{d-1} \mathrm{d}\Omega_d \, \langle |\mathbf{w}_{\delta}(\mathbf{k})|^2 \rangle,\quad \delta=\mathrm{v},\mathrm{q}.
\end{eqnarray}
Following Ref.~\cite{Nore1997a,Nore1997b}, the velocity $\mathbf{w}_{\mathrm{v}}$, 
is furthermore decomposed into `incompressible' (divergence free) and `compressible' (solenoidal) parts, $\mathbf{w}_{\mathrm{v}}=\mathbf{w}_{\mathrm{i}}+\mathbf{w}_{\mathrm{c}}$, with $\nabla \cdot \mathbf{w}_\mathrm{i}=0$, $\nabla \times \mathbf{w}_\mathrm{c}=0$, to distinguish vortical superfluid and rotationless motion of the fluid.


\begin{thebibliography}{77}
\expandafter\ifx\csname natexlab\endcsname\relax\def\natexlab#1{#1}\fi
\expandafter\ifx\csname bibnamefont\endcsname\relax
  \def\bibnamefont#1{#1}\fi
\expandafter\ifx\csname bibfnamefont\endcsname\relax
  \def\bibfnamefont#1{#1}\fi
\expandafter\ifx\csname citenamefont\endcsname\relax
  \def\citenamefont#1{#1}\fi
\expandafter\ifx\csname url\endcsname\relax
  \def\url#1{\texttt{#1}}\fi
\expandafter\ifx\csname urlprefix\endcsname\relax\def\urlprefix{URL }\fi
\providecommand{\bibinfo}[2]{#2}
\providecommand{\eprint}[2][]{\url{#2}}

\bibitem[{\citenamefont{Richardson}(1920)}]{Richardson1920a}
\bibinfo{author}{\bibfnamefont{L.~F.} \bibnamefont{Richardson}},
  \bibinfo{journal}{Proc. R. Soc. London. Ser. A}
  \textbf{\bibinfo{volume}{97}}, \bibinfo{pages}{354} (\bibinfo{year}{1920}).

\bibitem[{\citenamefont{Kolmogorov}(1941)}]{Kolmogorov1941a}
\bibinfo{author}{\bibfnamefont{A.~N.} \bibnamefont{Kolmogorov}},
  \bibinfo{journal}{Proc. USSR Acad. Sci.} \textbf{\bibinfo{volume}{30}},
  \bibinfo{pages}{{299}} (\bibinfo{year}{1941}), \bibinfo{note}{{[Proc. R. Soc. Lond. A 434, 9 (1991)]}}.

\bibitem[{\citenamefont{Obukhov}(1941)}]{Obukhov1941a}
\bibinfo{author}{\bibfnamefont{A.~M.} \bibnamefont{Obukhov}},
  \bibinfo{journal}{Izv. Akad. Nauk S.S.S.R., Ser. Geogr. Geofiz.}
  \textbf{\bibinfo{volume}{5}}, \bibinfo{pages}{{453}} (\bibinfo{year}{1941}).

\bibitem[{\citenamefont{Frisch}(1995)}]{Frisch1995a}
\bibinfo{author}{\bibfnamefont{U.}~\bibnamefont{Frisch}},
  \emph{\bibinfo{title}{Turbulence: The Legacy of A. N. Kolmogorov}}
  (\bibinfo{publisher}{Cambridge University Press, Cambridge, UK},
  \bibinfo{year}{1995}).

\bibitem[{\citenamefont{Nore et~al.}(1997{\natexlab{a}})\citenamefont{Nore,
  Abid, and Brachet}}]{Nore1997a}
\bibinfo{author}{\bibfnamefont{C.}~\bibnamefont{Nore}},
  \bibinfo{author}{\bibfnamefont{M.}~\bibnamefont{Abid}}, \bibnamefont{and}
  \bibinfo{author}{\bibfnamefont{M.~E.} \bibnamefont{Brachet}},
  \bibinfo{journal}{Phys. Rev. Lett.} \textbf{\bibinfo{volume}{78}},
  \bibinfo{pages}{3896} (\bibinfo{year}{1997}{\natexlab{a}}).

\bibitem[{\citenamefont{Vinen and Niemela}(2002)}]{Vinen2002a}
\bibinfo{author}{\bibfnamefont{W.~F.} \bibnamefont{Vinen}} \bibnamefont{and}
  \bibinfo{author}{\bibfnamefont{J.~J.} \bibnamefont{Niemela}},
  \bibinfo{journal}{J. Low Temp. Phys.} \textbf{\bibinfo{volume}{128}},
  \bibinfo{pages}{167} (\bibinfo{year}{2002}).

\bibitem[{\citenamefont{Araki et~al.}(2002)\citenamefont{Araki, Tsubota, and
  Nemirovskii}}]{Araki2002a}
\bibinfo{author}{\bibfnamefont{T.}~\bibnamefont{Araki}},
  \bibinfo{author}{\bibfnamefont{M.}~\bibnamefont{Tsubota}}, \bibnamefont{and}
  \bibinfo{author}{\bibfnamefont{S.~K.} \bibnamefont{Nemirovskii}},
  \bibinfo{journal}{Phys. Rev. Lett.} \textbf{\bibinfo{volume}{89}},
  \bibinfo{pages}{145301} (\bibinfo{year}{2002}).

\bibitem[{\citenamefont{Kobayashi and Tsubota}(2005)}]{Kobayashi2005a}
\bibinfo{author}{\bibfnamefont{M.}~\bibnamefont{Kobayashi}} \bibnamefont{and}
  \bibinfo{author}{\bibfnamefont{M.}~\bibnamefont{Tsubota}},
  \bibinfo{journal}{Phys. Rev. Lett.} \textbf{\bibinfo{volume}{94}},
  \bibinfo{pages}{065302} (\bibinfo{year}{2005}).

\bibitem[{\citenamefont{Proment et~al.}(2011)\citenamefont{Proment, Nazarenko,
  and Onorato}}]{Proment2010a}
\bibinfo{author}{\bibfnamefont{D.}~\bibnamefont{Proment}},
  \bibinfo{author}{\bibfnamefont{S.}~\bibnamefont{Nazarenko}},
  \bibnamefont{and} \bibinfo{author}{\bibfnamefont{M.}~\bibnamefont{Onorato}},
  \bibinfo{journal}{Physica D: Nonlin. Phen.} \textbf{\bibinfo{volume}{241}},
  \bibinfo{pages}{304} (\bibinfo{year}{2011}).

\bibitem[{\citenamefont{Henn et~al.}(2009)\citenamefont{Henn, Seman, Roati,
  {a}es, and Bagnato}}]{Henn2009a}
\bibinfo{author}{\bibfnamefont{E.~A.~L.} \bibnamefont{Henn}},  
  \bibnamefont{et~al.},
\bibinfo{journal}{Phys. Rev. Lett.}
  \textbf{\bibinfo{volume}{103}}, \bibinfo{eid}{045301}
  (\bibinfo{year}{2009}).

\bibitem[{\citenamefont{Seman et~al.}(2011)\citenamefont{Seman, Henn, Shiozaki,
  Roati, Poveda-Cuevas, Magalh\~{a}es, Yukalov, Tsubota, Kobayashi, Kasamatsu
  et~al.}}]{Seman2011a}
\bibinfo{author}{\bibfnamefont{J.}~\bibnamefont{Seman}},
  \bibnamefont{et~al.}, \bibinfo{journal}{Las. Phys. Lett.}
  \textbf{\bibinfo{volume}{8}}, \bibinfo{pages}{691} (\bibinfo{year}{2011}).

\bibitem[{\citenamefont{{Neely} et~al.}(2012)\citenamefont{{Neely}, {Bradley},
  {Samson}, {Rooney}, {Wright}, {Law}, {Carretero-Gonz{\'a}lez}, {Kevrekidis},
  {Davis}, and {Anderson}}}]{Neely2012a}
\bibinfo{author}{\bibfnamefont{T.~W.} \bibnamefont{{Neely}}},
  \bibnamefont{et~al.},
  \bibinfo{journal}{arXiv:1204.1102 [cond-mat.quant-gas]}
  (\bibinfo{year}{2012}).

\bibitem[{\citenamefont{Zakharov et~al.}(1992)\citenamefont{Zakharov, {L'vov},
  and Falkovich}}]{Zakharov1992a}
\bibinfo{author}{\bibfnamefont{V.~E.} \bibnamefont{Zakharov}},
  \bibinfo{author}{\bibfnamefont{V.~S.} \bibnamefont{{L'vov}}},
  \bibnamefont{and}
  \bibinfo{author}{\bibfnamefont{G.}~\bibnamefont{Falkovich}},
  \emph{\bibinfo{title}{Kolmogorov Spectra of Turbulence I: Wave Turbulence}}
  (\bibinfo{publisher}{Springer-Verlag, Berlin}, \bibinfo{year}{1992}).

\bibitem[{\citenamefont{Nazarenko}(2011)}]{Nazarenko2011a}
\bibinfo{author}{\bibfnamefont{S.}~\bibnamefont{Nazarenko}},
  \emph{\bibinfo{title}{Wave turbulence}}, no. \bibinfo{number}{825} in
  \bibinfo{series}{Lecture Notes in Physics} (\bibinfo{publisher}{Springer},
  \bibinfo{address}{Heidelberg etc.}, \bibinfo{year}{2011}).

\bibitem[{\citenamefont{Scheppach et~al.}(2010)\citenamefont{Scheppach, Berges,
  and Gasenzer}}]{Scheppach:2009wu}
\bibinfo{author}{\bibfnamefont{C.}~\bibnamefont{Scheppach}},
  \bibinfo{author}{\bibfnamefont{J.}~\bibnamefont{Berges}}, \bibnamefont{and}
  \bibinfo{author}{\bibfnamefont{T.}~\bibnamefont{Gasenzer}},
  \bibinfo{journal}{Phys. Rev. A} \textbf{\bibinfo{volume}{81}},
  \bibinfo{pages}{033611} (\bibinfo{year}{2010}).

\bibitem[{\citenamefont{Nowak et~al.}(2011{\natexlab{a}})\citenamefont{Nowak,
  Sexty, and Gasenzer}}]{Nowak:2010tm}
\bibinfo{author}{\bibfnamefont{B.}~\bibnamefont{Nowak}},
  \bibinfo{author}{\bibfnamefont{D.}~\bibnamefont{Sexty}}, \bibnamefont{and}
  \bibinfo{author}{\bibfnamefont{T.}~\bibnamefont{Gasenzer}},
  \bibinfo{journal}{Phys. Rev. B} \textbf{\bibinfo{volume}{84}},
  \bibinfo{pages}{020506(R)} (\bibinfo{year}{2011}{\natexlab{a}}).

\bibitem[{\citenamefont{Levich and Yakhot}(1978)}]{Levich1978a}
\bibinfo{author}{\bibfnamefont{E.}~\bibnamefont{Levich}} \bibnamefont{and}
  \bibinfo{author}{\bibfnamefont{V.}~\bibnamefont{Yakhot}},
  \bibinfo{journal}{J. Phys. A: Math. Gen.} \textbf{\bibinfo{volume}{11}},
  \bibinfo{pages}{2237} (\bibinfo{year}{1978}).

\bibitem[{\citenamefont{Kagan et~al.}(1992)\citenamefont{Kagan, Svistunov, and
  Shlyapnikov}}]{Kagan1992a}
\bibinfo{author}{\bibfnamefont{Y.}~\bibnamefont{Kagan}},
  \bibinfo{author}{\bibfnamefont{B.~V.} \bibnamefont{Svistunov}},
  \bibnamefont{and} \bibinfo{author}{\bibfnamefont{G.~V.}
  \bibnamefont{Shlyapnikov}}, \bibinfo{journal}{[Zh. Eksp. Teor. Fiz. 101, 528
  (1992)] Sov. Phys. JETP} \textbf{\bibinfo{volume}{74}}, \bibinfo{pages}{279}
  (\bibinfo{year}{1992}).

\bibitem[{\citenamefont{Kagan and Svistunov}(1994)}]{Kagan1994a}
\bibinfo{author}{\bibfnamefont{Y.}~\bibnamefont{Kagan}} \bibnamefont{and}
  \bibinfo{author}{\bibfnamefont{B.~V.} \bibnamefont{Svistunov}},
  \bibinfo{journal}{[Zh. Eksp. Teor. Fiz. 105, 353 (1994)] Sov. Phys. JETP}
  \textbf{\bibinfo{volume}{78}}, \bibinfo{pages}{187} (\bibinfo{year}{1994}).

\bibitem[{\citenamefont{Berloff and Svistunov}(2002)}]{Berloff2002a}
\bibinfo{author}{\bibfnamefont{N.~G.} \bibnamefont{Berloff}} \bibnamefont{and}
  \bibinfo{author}{\bibfnamefont{B.~V.} \bibnamefont{Svistunov}},
  \bibinfo{journal}{Phys. Rev. A} \textbf{\bibinfo{volume}{66}},
  \bibinfo{pages}{013603} (\bibinfo{year}{2002}).

\bibitem[{\citenamefont{Svistunov}(2001)}]{Svistunov2001a}
\bibinfo{author}{\bibfnamefont{B.}~\bibnamefont{Svistunov}}, in
  \emph{\bibinfo{booktitle}{Quantized Vortex Dynamics and Superfluid
  Turbulence}}, edited by
  \bibinfo{editor}{\bibfnamefont{C.}~\bibnamefont{Barenghi}},
  \bibinfo{editor}{\bibfnamefont{R.}~\bibnamefont{Donnelly}}, \bibnamefont{and}
  \bibinfo{editor}{\bibfnamefont{W.}~\bibnamefont{Vinen}}
  (\bibinfo{publisher}{Springer, Berlin etc.}, \bibinfo{year}{2001}), no.
  \bibinfo{number}{571} in \bibinfo{series}{Lecture Notes in Physics}.

\bibitem[{\citenamefont{Kozik and Svistunov}(2009)}]{Kozik2009a}
\bibinfo{author}{\bibfnamefont{E.~V.} \bibnamefont{Kozik}} \bibnamefont{and}
  \bibinfo{author}{\bibfnamefont{B.~V.} \bibnamefont{Svistunov}},
  \bibinfo{journal}{J. Low Temp. Phys.} \textbf{\bibinfo{volume}{156}},
  \bibinfo{pages}{215} (\bibinfo{year}{2009}).

\bibitem[{\citenamefont{Kagan and Svistunov}(1997)}]{Kagan1997c}
\bibinfo{author}{\bibfnamefont{Y.}~\bibnamefont{Kagan}} \bibnamefont{and}
  \bibinfo{author}{\bibfnamefont{B.~V.} \bibnamefont{Svistunov}},
  \bibinfo{journal}{Phys. Rev. Lett.} \textbf{\bibinfo{volume}{79}},
  \bibinfo{pages}{3331} (\bibinfo{year}{1997}).

\bibitem[{\citenamefont{Nowak et~al.}(2011{\natexlab{b}})\citenamefont{Nowak,
  Schole, Sexty, and Gasenzer}}]{Nowak:2011sk}
\bibinfo{author}{\bibfnamefont{B.}~\bibnamefont{Nowak}},
  \bibinfo{author}{\bibfnamefont{J.}~\bibnamefont{Schole}},
  \bibinfo{author}{\bibfnamefont{D.}~\bibnamefont{Sexty}}, \bibnamefont{and}
  \bibinfo{author}{\bibfnamefont{T.}~\bibnamefont{Gasenzer}},
  \bibinfo{journal}{Phys. Rev. A} \textbf{\bibinfo{volume}{85}},
  \bibinfo{pages}{043627} (\bibinfo{year}{2012}{\natexlab{b}}).

\bibitem[{\citenamefont{Schmidt et~al.}(2012)\citenamefont{Schmidt, Erne,
  Nowak, Sexty, and Gasenzer}}]{Schmidt:2012kw}
\bibinfo{author}{\bibfnamefont{M.}~\bibnamefont{Schmidt}},
  \bibinfo{author}{\bibfnamefont{S.}~\bibnamefont{Erne}},
  \bibinfo{author}{\bibfnamefont{B.}~\bibnamefont{Nowak}},
  \bibinfo{author}{\bibfnamefont{D.}~\bibnamefont{Sexty}}, \bibnamefont{and}
  \bibinfo{author}{\bibfnamefont{T.}~\bibnamefont{Gasenzer}},
  \bibinfo{journal}{arXiv:1203.3651 [cond-mat.quant-gas]}
New J. Phys., to appear
  (\bibinfo{year}{2012}).

\bibitem[{\citenamefont{Micha and Tkachev}(2003)}]{Micha:2002ey}
\bibinfo{author}{\bibfnamefont{R.}~\bibnamefont{Micha}} \bibnamefont{and}
  \bibinfo{author}{\bibfnamefont{I.~I.} \bibnamefont{Tkachev}},
  \bibinfo{journal}{Phys. Rev. Lett.} \textbf{\bibinfo{volume}{90}},
  \bibinfo{pages}{121301} (\bibinfo{year}{2003}).

\bibitem[{\citenamefont{Berges et~al.}(2008)\citenamefont{Berges, Rothkopf, and
  Schmidt}}]{Berges:2008wm}
\bibinfo{author}{\bibfnamefont{J.}~\bibnamefont{Berges}},
  \bibinfo{author}{\bibfnamefont{A.}~\bibnamefont{Rothkopf}}, \bibnamefont{and}
  \bibinfo{author}{\bibfnamefont{J.}~\bibnamefont{Schmidt}},
  \bibinfo{journal}{Phys. Rev. Lett.} \textbf{\bibinfo{volume}{101}},
  \bibinfo{pages}{041603} (\bibinfo{year}{2008}).

\bibitem[{\citenamefont{Berges and Hoffmeister}(2009)}]{Berges:2008sr}
\bibinfo{author}{\bibfnamefont{J.}~\bibnamefont{Berges}} \bibnamefont{and}
  \bibinfo{author}{\bibfnamefont{G.}~\bibnamefont{Hoffmeister}},
  \bibinfo{journal}{Nucl. Phys.} \textbf{\bibinfo{volume}{B813}},
  \bibinfo{pages}{383} (\bibinfo{year}{2009}).

\bibitem[{\citenamefont{Berges and Sexty}(2011)}]{Berges:2010ez}
\bibinfo{author}{\bibfnamefont{J.}~\bibnamefont{Berges}} \bibnamefont{and}
  \bibinfo{author}{\bibfnamefont{D.}~\bibnamefont{Sexty}},
  \bibinfo{journal}{Phys. Rev. D} \textbf{\bibinfo{volume}{83}},
  \bibinfo{pages}{085004} (\bibinfo{year}{2011}).

\bibitem[{\citenamefont{Gasenzer et~al.}(2011)\citenamefont{Gasenzer, Nowak,
  and Sexty}}]{Gasenzer:2011by}
\bibinfo{author}{\bibfnamefont{T.}~\bibnamefont{Gasenzer}},
  \bibinfo{author}{\bibfnamefont{B.}~\bibnamefont{Nowak}}, \bibnamefont{and}
  \bibinfo{author}{\bibfnamefont{D.}~\bibnamefont{Sexty}},
  \bibinfo{journal}{Phys. Lett. B} \textbf{\bibinfo{volume}{710}},
  \bibinfo{pages}{500} (\bibinfo{year}{2012}).

\bibitem[{\citenamefont{Arnold and Moore}(2006{\natexlab{a}})}]{Arnold:2005ef}
\bibinfo{author}{\bibfnamefont{P.~B.} \bibnamefont{Arnold}} \bibnamefont{and}
  \bibinfo{author}{\bibfnamefont{G.~D.} \bibnamefont{Moore}},
  \bibinfo{journal}{Phys. Rev.} \textbf{\bibinfo{volume}{D73}},
  \bibinfo{pages}{025006} (\bibinfo{year}{2006}{\natexlab{a}});
%
  \bibinfo{journal}{Phys. Rev.} \textbf{\bibinfo{volume}{D73}},
  \bibinfo{pages}{025013} (\bibinfo{year}{2006}{\natexlab{b}}).

\bibitem[{\citenamefont{Berges et~al.}(2009)\citenamefont{Berges, Scheffler,
  and Sexty}}]{Berges:2008mr}
\bibinfo{author}{\bibfnamefont{J.}~\bibnamefont{Berges}},
  \bibinfo{author}{\bibfnamefont{S.}~\bibnamefont{Scheffler}},
  \bibnamefont{and} \bibinfo{author}{\bibfnamefont{D.}~\bibnamefont{Sexty}},
  \bibinfo{journal}{Phys. Lett.} \textbf{\bibinfo{volume}{B681}},
  \bibinfo{pages}{362} (\bibinfo{year}{2009}).

\bibitem[{\citenamefont{Carrington and Rebhan}(2011)}]{Carrington:2010sz}
\bibinfo{author}{\bibfnamefont{M.}~\bibnamefont{Carrington}} \bibnamefont{and}
  \bibinfo{author}{\bibfnamefont{A.}~\bibnamefont{Rebhan}},
  \bibinfo{journal}{Eur. Phys. J.} \textbf{\bibinfo{volume}{C71}},
  \bibinfo{pages}{1787} (\bibinfo{year}{2011}).

\bibitem[{\citenamefont{Blaizot et~al.}(2012)\citenamefont{Blaizot, Gelis,
  Liao, McLerran, and Venugopalan}}]{Blaizot:2011xf}
\bibinfo{author}{\bibfnamefont{J.-P.} \bibnamefont{Blaizot}},
  \bibinfo{author}{\bibfnamefont{F.}~\bibnamefont{Gelis}},
  \bibinfo{author}{\bibfnamefont{J.-F.} \bibnamefont{Liao}},
  \bibinfo{author}{\bibfnamefont{L.}~\bibnamefont{McLerran}}, \bibnamefont{and}
  \bibinfo{author}{\bibfnamefont{R.}~\bibnamefont{Venugopalan}},
  \bibinfo{journal}{Nucl.Phys.} \textbf{\bibinfo{volume}{A873}},
  \bibinfo{pages}{68} (\bibinfo{year}{2012}).

\bibitem[{\citenamefont{Berges and Sexty}(2012)}]{Berges:2012us}
\bibinfo{author}{\bibfnamefont{J.}~\bibnamefont{Berges}} \bibnamefont{and}
  \bibinfo{author}{\bibfnamefont{D.}~\bibnamefont{Sexty}}, 
\bibinfo{journal}{Phys. Rev. Lett.} \textbf{\bibinfo{volume}{108}},
  \bibinfo{pages}{161601} (\bibinfo{year}{2012}).

\bibitem[{\citenamefont{Fukushima and Gelis}(2012)}]{Fukushima:2011nq}
\bibinfo{author}{\bibfnamefont{K.}~\bibnamefont{Fukushima}} \bibnamefont{and}
  \bibinfo{author}{\bibfnamefont{F.}~\bibnamefont{Gelis}},
  \bibinfo{journal}{Nucl.Phys.} \textbf{\bibinfo{volume}{A874}},
  \bibinfo{pages}{108} (\bibinfo{year}{2012}).

\bibitem[{\citenamefont{Fukushima}(2011)}]{Fukushima:2011ca}
\bibinfo{author}{\bibfnamefont{K.}~\bibnamefont{Fukushima}},
  \bibinfo{journal}{Acta Phys. Polon.} \textbf{\bibinfo{volume}{B42}},
  \bibinfo{pages}{2697} (\bibinfo{year}{2011}).

\bibitem[{\citenamefont{Hohenberg and Halperin}(1977)}]{Hohenberg1977a}
\bibinfo{author}{\bibfnamefont{P.~C.} \bibnamefont{Hohenberg}}
  \bibnamefont{and} \bibinfo{author}{\bibfnamefont{B.~I.}
  \bibnamefont{Halperin}}, \bibinfo{journal}{Rev. Mod. Phys.}
  \textbf{\bibinfo{volume}{49}}, \bibinfo{pages}{435} (\bibinfo{year}{1977}).

\bibitem[{\citenamefont{Timm}(1996)}]{Timm1996a}
\bibinfo{author}{\bibfnamefont{C.}~\bibnamefont{Timm}},
  \bibinfo{journal}{Physica C: Superconductivity}
  \textbf{\bibinfo{volume}{265}}, \bibinfo{pages}{31} (\bibinfo{year}{1996}).

\bibitem[{\citenamefont{Bisset et~al.}(2009)\citenamefont{Bisset, Davis,
  Simula, and Blakie}}]{Bisset2009a}
\bibinfo{author}{\bibfnamefont{R.~N.} \bibnamefont{Bisset}},
  \bibinfo{author}{\bibfnamefont{M.~J.} \bibnamefont{Davis}},
  \bibinfo{author}{\bibfnamefont{T.~P.} \bibnamefont{Simula}},
  \bibnamefont{and} \bibinfo{author}{\bibfnamefont{P.~B.}
  \bibnamefont{Blakie}}, \bibinfo{journal}{Phys. Rev. A}
  \textbf{\bibinfo{volume}{79}}, \bibinfo{pages}{033626}
  (\bibinfo{year}{2009}).

\bibitem[{\citenamefont{Foster et~al.}(2010)\citenamefont{Foster, Blakie, and
  Davis}}]{Foster2010a}
\bibinfo{author}{\bibfnamefont{C.~J.} \bibnamefont{Foster}},
  \bibinfo{author}{\bibfnamefont{P.~B.} \bibnamefont{Blakie}},
  \bibnamefont{and} \bibinfo{author}{\bibfnamefont{M.~J.} \bibnamefont{Davis}},
  \bibinfo{journal}{Phys. Rev. A} \textbf{\bibinfo{volume}{81}},
  \bibinfo{pages}{023623} (\bibinfo{year}{2010}).

\bibitem[{\citenamefont{Damle et~al.}(1996)\citenamefont{Damle, Majumdar, and
  Sachdev}}]{Damle1996a}
\bibinfo{author}{\bibfnamefont{K.}~\bibnamefont{Damle}},
  \bibinfo{author}{\bibfnamefont{S.~N.}~\bibnamefont{Majumdar}}, \bibnamefont{and}
  \bibinfo{author}{\bibfnamefont{S.}~\bibnamefont{Sachdev}},
  \bibinfo{journal}{Phys. Rev. A} \textbf{\bibinfo{volume}{54}},
  \bibinfo{pages}{5037} (\bibinfo{year}{1996}).

\bibitem[{\citenamefont{Nazarenko and Onorato}(2006)}]{Nazarenko2006a}
\bibinfo{author}{\bibfnamefont{S.}~\bibnamefont{Nazarenko}} \bibnamefont{and}
  \bibinfo{author}{\bibfnamefont{M.}~\bibnamefont{Onorato}},
  \bibinfo{journal}{Physica D: Nonlin. Phen.} \textbf{\bibinfo{volume}{219}},
  \bibinfo{pages}{1} (\bibinfo{year}{2006}).

\bibitem[{\citenamefont{Onsager}(1949)}]{Onsager1949a}
\bibinfo{author}{\bibfnamefont{L.}~\bibnamefont{Onsager}},
  \bibinfo{journal}{Nuovo Cim. Suppl.} \textbf{\bibinfo{volume}{6}},
  \bibinfo{pages}{279} (\bibinfo{year}{1949}).

\bibitem[{\citenamefont{Berezinskii}(1971)}]{Berezinskii1971a}
\bibinfo{author}{\bibfnamefont{V.}~\bibnamefont{Berezinskii}},
  \bibinfo{journal}{Sov. J. Exp. Theor. Phys.}
  \textbf{\bibinfo{volume}{32}}, \bibinfo{pages}{493} (\bibinfo{year}{1971}).

\bibitem[{\citenamefont{Kosterlitz and Thouless}(1973)}]{Kosterlitz1973a}
\bibinfo{author}{\bibfnamefont{J.}~\bibnamefont{Kosterlitz}} \bibnamefont{and}
  \bibinfo{author}{\bibfnamefont{D.}~\bibnamefont{Thouless}},
  \bibinfo{journal}{J. Phys. C: Solid State Phys.}
  \textbf{\bibinfo{volume}{6}}, \bibinfo{pages}{1181} (\bibinfo{year}{1973}).

\bibitem[{\citenamefont{Kibble}(1976)}]{Kibble1976a}
\bibinfo{author}{\bibfnamefont{T.~W.~B.} \bibnamefont{Kibble}},
  \bibinfo{journal}{J. Phys. A: Math. Gen.}
  \textbf{\bibinfo{volume}{9}}, \bibinfo{pages}{1387} (\bibinfo{year}{1976}).

\bibitem[{\citenamefont{Zurek}(1985)}]{Zurek1985a}
\bibinfo{author}{\bibfnamefont{W.~H.} \bibnamefont{Zurek}},
  \bibinfo{journal}{Nature} \textbf{\bibinfo{volume}{317}},
  \bibinfo{pages}{505} (\bibinfo{year}{1985}).

\bibitem[{\citenamefont{Mathey et~al.}(2011)\citenamefont{Mathey, G{\"u}nter,
  Dalibard, and Polkovnikov}}]{Mathey2012a}
\bibinfo{author}{\bibfnamefont{L.}~\bibnamefont{Mathey}},
  \bibinfo{author}{\bibfnamefont{K.}~\bibnamefont{G{\"u}nter}},
  \bibinfo{author}{\bibfnamefont{J.}~\bibnamefont{Dalibard}}, \bibnamefont{and}
  \bibinfo{author}{\bibfnamefont{A.}~\bibnamefont{Polkovnikov}},
  \bibinfo{journal}{arXiv:1112.1204 [cond-mat.quant-gas]}
  (\bibinfo{year}{2011}).

\bibitem[{\citenamefont{Weiler et~al.}(2008)\citenamefont{Weiler, Neely,
  Scherer, Bradley, Davis, and Anderson}}]{Weiler2008a}
\bibinfo{author}{\bibfnamefont{C.~N.} \bibnamefont{Weiler}},
  \bibinfo{author}{\bibfnamefont{T.~W.} \bibnamefont{Neely}},
  \bibinfo{author}{\bibfnamefont{D.~R.} \bibnamefont{Scherer}},
  \bibinfo{author}{\bibfnamefont{A.~S.} \bibnamefont{Bradley}},
  \bibinfo{author}{\bibfnamefont{M.~J.} \bibnamefont{Davis}}, \bibnamefont{and}
  \bibinfo{author}{\bibfnamefont{B.~P.} \bibnamefont{Anderson}},
  \bibinfo{journal}{Nature} \textbf{\bibinfo{volume}{455}},
  \bibinfo{pages}{948} (\bibinfo{year}{2008}).

\bibitem[{\citenamefont{Eyink and Goldenfeld}(1994)}]{Eyink1994a}
\bibinfo{author}{\bibfnamefont{G.}~\bibnamefont{Eyink}} \bibnamefont{and}
  \bibinfo{author}{\bibfnamefont{N.}~\bibnamefont{Goldenfeld}},
  \bibinfo{journal}{Phys. Rev. E} \textbf{\bibinfo{volume}{50}},
  \bibinfo{pages}{4679} (\bibinfo{year}{1994}).

\bibitem[{\citenamefont{Mathey and Polkovnikov}(2009)}]{Mathey2009a}
\bibinfo{author}{\bibfnamefont{L.}~\bibnamefont{Mathey}} \bibnamefont{and}
  \bibinfo{author}{\bibfnamefont{A.}~\bibnamefont{Polkovnikov}},
  \bibinfo{journal}{Phys. Rev. A} \textbf{\bibinfo{volume}{80}},
  \bibinfo{pages}{041601} (\bibinfo{year}{2009});
%
  \bibinfo{journal}{Phys. Rev. A} \textbf{\bibinfo{volume}{81}},
  \bibinfo{pages}{033605} (\bibinfo{year}{2010}).

\bibitem[{\citenamefont{Mitra and Giamarchi}(2011)}]{Mitra2011a}
\bibinfo{author}{\bibfnamefont{A.}~\bibnamefont{Mitra}} \bibnamefont{and}
  \bibinfo{author}{\bibfnamefont{T.}~\bibnamefont{Giamarchi}},
  \bibinfo{journal}{Phys. Rev. Lett.} \textbf{\bibinfo{volume}{107}},
  \bibinfo{pages}{150602} (\bibinfo{year}{2011});
%
  \bibinfo{journal}{Phys. Rev. B} \textbf{\bibinfo{volume}{85}},
  \bibinfo{pages}{075117} (\bibinfo{year}{2012}).

\bibitem[{\citenamefont{Canet and Chate}(2007)}]{Canet:2006xu}
\bibinfo{author}{\bibfnamefont{L.}~\bibnamefont{Canet}} \bibnamefont{and}
  \bibinfo{author}{\bibfnamefont{H.}~\bibnamefont{Chate}}, \bibinfo{journal}{J.
  Phys. A} \textbf{\bibinfo{volume}{40}}, \bibinfo{pages}{1937}
  (\bibinfo{year}{2007}).

\bibitem[{\citenamefont{Gasenzer and Pawlowski}(2008)}]{Gasenzer:2008zz}
\bibinfo{author}{\bibfnamefont{T.}~\bibnamefont{Gasenzer}} \bibnamefont{and}
  \bibinfo{author}{\bibfnamefont{J.~M.} \bibnamefont{Pawlowski}},
  \bibinfo{journal}{Phys. Lett.} \textbf{\bibinfo{volume}{B670}},
  \bibinfo{pages}{135} (\bibinfo{year}{2008}).

\bibitem[{\citenamefont{Gasenzer et~al.}(2010)\citenamefont{Gasenzer, Kessler,
  and Pawlowski}}]{Gasenzer:2010rq}
\bibinfo{author}{\bibfnamefont{T.}~\bibnamefont{Gasenzer}},
  \bibinfo{author}{\bibfnamefont{S.}~\bibnamefont{Kessler}}, \bibnamefont{and}
  \bibinfo{author}{\bibfnamefont{J.~M.} \bibnamefont{Pawlowski}},
  \bibinfo{journal}{Eur. Phys. J. C} \textbf{\bibinfo{volume}{70}},
  \bibinfo{pages}{423} (\bibinfo{year}{2010}).

\bibitem[{\citenamefont{videos}(2011)}]{videos}\bibinfo{author}{For videos of the evolution see: http://www.thphys.uni-heidelberg.de/{$\sim$}smp/gasenzer/videos/boseqt.html}

\bibitem[{\citenamefont{Aref}(1983)}]{Aref1983a}
\bibinfo{author}{\bibfnamefont{H.}~\bibnamefont{Aref}}, \bibinfo{journal}{Ann.
  Rev. Fl. Mech.} \textbf{\bibinfo{volume}{15}}, \bibinfo{pages}{345}
  (\bibinfo{year}{1983}).

\bibitem[{\citenamefont{Nore et~al.}(1997{\natexlab{b}})\citenamefont{Nore,
  Abid, and Brachet}}]{Nore1997b}
\bibinfo{author}{\bibfnamefont{C.}~\bibnamefont{Nore}},
  \bibinfo{author}{\bibfnamefont{M.}~\bibnamefont{Abid}}, \bibnamefont{and}
  \bibinfo{author}{\bibfnamefont{M.~E.} \bibnamefont{Brachet}},
  \bibinfo{journal}{Phys. Fluids} \textbf{\bibinfo{volume}{9}},
  \bibinfo{pages}{2644} (\bibinfo{year}{1997}{\natexlab{b}}).

\bibitem[{\citenamefont{Krstulovic and Brachet}(2011)}]{Krstulovic2011a}
\bibinfo{author}{\bibfnamefont{G.}~\bibnamefont{Krstulovic}} \bibnamefont{and}
  \bibinfo{author}{\bibfnamefont{M.}~\bibnamefont{Brachet}},
  \bibinfo{journal}{Phys. Rev. E} \textbf{\bibinfo{volume}{83}},
  \bibinfo{pages}{066311} (\bibinfo{year}{2011}).

\bibitem[{\citenamefont{Ambegaokar et~al.}(1980)}]{Ambegaokar1980a}
\bibinfo{author}{\bibfnamefont{V.}~\bibnamefont{Ambegaokar}}, 
  \bibinfo{author}{\bibfnamefont{B.~I.}~\bibnamefont{Halperin}},
\bibinfo{author}{\bibfnamefont{D.~R.}~\bibnamefont{Nelson}} \bibnamefont{and}
\bibinfo{author}{\bibfnamefont{E.~D.}~\bibnamefont{Siggia}},
  \bibinfo{journal}{Phys. Rev. B} \textbf{\bibinfo{volume}{21}},
  \bibinfo{pages}{1806} (\bibinfo{year}{1980}).

\bibitem[{\citenamefont{Carnevale et~al.}(1991)\citenamefont{Carnevale,
  McWilliams, Pomeau, Weiss, and Young}}]{Carnevale1991a}
\bibinfo{author}{\bibfnamefont{G.~F.} \bibnamefont{Carnevale}},
  \bibinfo{author}{\bibfnamefont{J.~C.} \bibnamefont{McWilliams}},
  \bibinfo{author}{\bibfnamefont{Y.}~\bibnamefont{Pomeau}},
  \bibinfo{author}{\bibfnamefont{J.~B.} \bibnamefont{Weiss}}, \bibnamefont{and}
  \bibinfo{author}{\bibfnamefont{W.~R.} \bibnamefont{Young}},
  \bibinfo{journal}{Phys. Rev. Lett.} \textbf{\bibinfo{volume}{66}},
  \bibinfo{pages}{2735} (\bibinfo{year}{1991}).

\bibitem[{\citenamefont{Weiss and McWilliams}(1993)}]{Weiss1993a}
\bibinfo{author}{\bibfnamefont{J.~B.} \bibnamefont{Weiss}} \bibnamefont{and}
  \bibinfo{author}{\bibfnamefont{J.~C.} \bibnamefont{McWilliams}},
  \bibinfo{journal}{Phys. Fluids A} \textbf{\bibinfo{volume}{3}},
  \bibinfo{pages}{608} (\bibinfo{year}{1993}).

\bibitem[{\citenamefont{Bracco et~al.}(2000)\citenamefont{Bracco, McWilliams,
  Murante, Provenzale, and Weiss}}]{Bracco2000a}
\bibinfo{author}{\bibfnamefont{A.}~\bibnamefont{Bracco}},
  \bibinfo{author}{\bibfnamefont{J.~C.} \bibnamefont{McWilliams}},
  \bibinfo{author}{\bibfnamefont{G.}~\bibnamefont{Murante}},
  \bibinfo{author}{\bibfnamefont{A.}~\bibnamefont{Provenzale}},
  \bibnamefont{and} \bibinfo{author}{\bibfnamefont{J.~B.} \bibnamefont{Weiss}},
  \bibinfo{journal}{Phys. Fluids} \textbf{\bibinfo{volume}{12}},
  \bibinfo{pages}{2931} (\bibinfo{year}{2000}).

\bibitem[{\citenamefont{Tabeling et~al.}(1991)\citenamefont{Tabeling, Burkhart,
  Cardoso, and Willaime}}]{Tabeling:decay:1991}
\bibinfo{author}{\bibfnamefont{P.}~\bibnamefont{Tabeling}},
  \bibinfo{author}{\bibfnamefont{S.}~\bibnamefont{Burkhart}},
  \bibinfo{author}{\bibfnamefont{O.}~\bibnamefont{Cardoso}}, \bibnamefont{and}
  \bibinfo{author}{\bibfnamefont{H.}~\bibnamefont{Willaime}},
  \bibinfo{journal}{Phys. Rev. Lett.} \textbf{\bibinfo{volume}{67}},
  \bibinfo{pages}{3772} (\bibinfo{year}{1991}).

\bibitem[{\citenamefont{Benzi et~al.}(1992)\citenamefont{Benzi, Colella,
  Briscolini, and Santangelo}}]{Benzi:decay:1992}
\bibinfo{author}{\bibfnamefont{R.}~\bibnamefont{Benzi}},
  \bibinfo{author}{\bibfnamefont{M.}~\bibnamefont{Colella}},
  \bibinfo{author}{\bibfnamefont{M.}~\bibnamefont{Briscolini}},
  \bibnamefont{and}
  \bibinfo{author}{\bibfnamefont{P.}~\bibnamefont{Santangelo}},
  \bibinfo{journal}{Phys. Fluids A} \textbf{\bibinfo{volume}{4}},
  \bibinfo{pages}{1036} (\bibinfo{year}{1992}).

\bibitem[{\citenamefont{Sire et~al.}(2011)\citenamefont{Sire, Chavanis, and
  Sopik}}]{Sire:merging:2010}
\bibinfo{author}{\bibfnamefont{C.}~\bibnamefont{Sire}},
  \bibinfo{author}{\bibfnamefont{P.-H.} \bibnamefont{Chavanis}},
  \bibnamefont{and} \bibinfo{author}{\bibfnamefont{J.}~\bibnamefont{Sopik}},
  \bibinfo{journal}{Phys. Rev. E} \textbf{\bibinfo{volume}{84}},
  \bibinfo{pages}{056317} (\bibinfo{year}{2011}).

\bibitem[{\citenamefont{Sire and Chavanis}(2000)}]{Sire:threebody:2000}
\bibinfo{author}{\bibfnamefont{C.}~\bibnamefont{Sire}} \bibnamefont{and}
  \bibinfo{author}{\bibfnamefont{P.-H.} \bibnamefont{Chavanis}},
  \bibinfo{journal}{Phys. Rev. E} \textbf{\bibinfo{volume}{61}},
  \bibinfo{pages}{6644} (\bibinfo{year}{2000}).

\bibitem[{\citenamefont{Yakhot and Wanderer}(2004)}]{Yakhot:crossover:2004}
\bibinfo{author}{\bibfnamefont{V.}~\bibnamefont{Yakhot}} \bibnamefont{and}
  \bibinfo{author}{\bibfnamefont{J.}~\bibnamefont{Wanderer}},
  \bibinfo{journal}{Phys. Rev. Lett.} \textbf{\bibinfo{volume}{93}},
  \bibinfo{pages}{154502} (\bibinfo{year}{2004}).

\bibitem[{\citenamefont{Chu and Williams}(2001)}]{Chu2001a}
\bibinfo{author}{\bibfnamefont{H.-C.}~\bibnamefont{Chu}} \bibnamefont{and}
\bibinfo{author}{\bibfnamefont{G.~A.}~\bibnamefont{Williams}},
  \bibinfo{journal}{Phys. Rev. Lett.} \textbf{\bibinfo{volume}{86}},
  \bibinfo{pages}{2585} (\bibinfo{year}{2001}).

\bibitem[{\citenamefont{Nazarenko and Onorato}(2007)}]{Nazarenko2007a}
\bibinfo{author}{\bibfnamefont{S.}~\bibnamefont{Nazarenko}} \bibnamefont{and}
  \bibinfo{author}{\bibfnamefont{M.}~\bibnamefont{Onorato}},
  \bibinfo{journal}{J. Low Temp. Phys.} \textbf{\bibinfo{volume}{146}},
  \bibinfo{pages}{31} (\bibinfo{year}{2007}).

\bibitem[{\citenamefont{Numasato et~al.}(2010)\citenamefont{Numasato, Tsubota,
  and L{'}vov}}]{Numasato2010a}
\bibinfo{author}{\bibfnamefont{R.}~\bibnamefont{Numasato}},
  \bibinfo{author}{\bibfnamefont{M.}~\bibnamefont{Tsubota}}, \bibnamefont{and}
  \bibinfo{author}{\bibfnamefont{V.~S.}~\bibnamefont{L{'}vov}},
  \bibinfo{journal}{Phys. Rev. A} \textbf{\bibinfo{volume}{81}},
  \bibinfo{pages}{063630} (\bibinfo{year}{2010}).

\bibitem[{\citenamefont{Bray}(1994)}]{Bray1994a}
\bibinfo{author}{\bibfnamefont{A.~J.} \bibnamefont{Bray}},
  \bibinfo{journal}{Adv. Phys.} \textbf{\bibinfo{volume}{43}},
  \bibinfo{pages}{357} (\bibinfo{year}{1994}).

\bibitem[{\citenamefont{Hadzibabic et~al.}(2006)\citenamefont{Hadzibabic,
  Kr{\"u}ger, Cheneau, Battelier, and Dalibard}}]{Hadzibabic2006a}
\bibinfo{author}{\bibfnamefont{Z.}~\bibnamefont{Hadzibabic}},
  \bibinfo{author}{\bibfnamefont{P.}~\bibnamefont{Kr{\"u}ger}},
  \bibinfo{author}{\bibfnamefont{M.}~\bibnamefont{Cheneau}},
  \bibinfo{author}{\bibfnamefont{B.}~\bibnamefont{Battelier}},
  \bibnamefont{and} \bibinfo{author}{\bibfnamefont{J.}~\bibnamefont{Dalibard}},
  \bibinfo{journal}{Nature} \textbf{\bibinfo{volume}{441}},
  \bibinfo{pages}{1118} (\bibinfo{year}{2006}).

\bibitem[{\citenamefont{Simula and Blakie}(2006)}]{Simula2006a}
\bibinfo{author}{\bibfnamefont{T.~P.} \bibnamefont{Simula}} \bibnamefont{and}
  \bibinfo{author}{\bibfnamefont{P.~B.} \bibnamefont{Blakie}},
  \bibinfo{journal}{Phys. Rev. Lett.} \textbf{\bibinfo{volume}{96}},
  \bibinfo{pages}{020404} (\bibinfo{year}{2006}).

\bibitem[{\citenamefont{Schweikhard et~al.}(2007)\citenamefont{Schweikhard,
  Tung, and Cornell}}]{Schweikhard2007a}
\bibinfo{author}{\bibfnamefont{V.}~\bibnamefont{Schweikhard}},
  \bibinfo{author}{\bibfnamefont{S.}~\bibnamefont{Tung}}, \bibnamefont{and}
  \bibinfo{author}{\bibfnamefont{E.~A.} \bibnamefont{Cornell}},
  \bibinfo{journal}{Phys. Rev. Lett.} \textbf{\bibinfo{volume}{99}},
  \bibinfo{pages}{030401} (\bibinfo{year}{2007}).

\bibitem[{\citenamefont{Giorgetti et~al.}(2007)\citenamefont{Giorgetti,
  Carusotto, and Castin}}]{Giorgetti2007a}
\bibinfo{author}{\bibfnamefont{L.}~\bibnamefont{Giorgetti}},
  \bibinfo{author}{\bibfnamefont{I.}~\bibnamefont{Carusotto}},
  \bibnamefont{and} \bibinfo{author}{\bibfnamefont{Y.}~\bibnamefont{Castin}},
  \bibinfo{journal}{Phys. Rev. A} \textbf{\bibinfo{volume}{76}},
  \bibinfo{pages}{013613} (\bibinfo{year}{2007}).

\bibitem[{\citenamefont{Kramer and MacKinnon}(1993)}]{Kramer1993a}
\bibinfo{author}{\bibfnamefont{B.}~\bibnamefont{Kramer}} \bibnamefont{and}
  \bibinfo{author}{\bibfnamefont{A.}~\bibnamefont{MacKinnon}},
  \bibinfo{journal}{Rep. Progr. Phys.} \textbf{\bibinfo{volume}{56}},
  \bibinfo{pages}{1469} (\bibinfo{year}{1993}).

\end{thebibliography}
\end{document}